# HATS-60b–HATS-69b: TEN TRANSITING PLANETS FROM HATSOUTH *

J. D. Hartman,[1] G. Á. Bakos,[1,2,†] D. Bayliss,[3] J. Bento,[4] W. Bhatti,[1] R. Brahm,[5,6,7] Z. Csubry,[1] N. Espinoza,[8,‡] Th. Henning,[8] A. Jordán,[7,6] L. Mancini,[9,8,10] K. Penev,[11] M. Rabus,[6,8] P. Sarkis,[8] V. Suc,[6] M. de Val-Borro,[12] G. Zhou,[13] B. Addison,[14] P. Arriagada,[15] R. P. Butler,[15] J. Crane,[16] S. Durkan,[17] S. Shectman,[16] T. G. Tan,[18] I. Thompson,[16] C. G. Tinney,[19,20] D. J. Wright,[19,20] J. Lázár,[21] I. Papp,[21] and P. Sári[21]

[1] *Department of Astrophysical Sciences, Princeton University, NJ 08544, USA*
[2] *MTA Distinguished Guest Fellow, Konkoly Observatory, Hungary*
[3] *Department of Physics, University of Warwick, Coventry CV4 7AL, UK*
[4] *Research School of Astronomy and Astrophysics, Australian National University, Canberra, ACT 2611, Australia*
[5] *Center of Astro-Engineering UC, Pontificia Universidad Católica de Chile, Av. Vicuña Mackenna 4860, 7820436 Macul, Santiago, Chile*
[6] *Instituto de Astrofísica, Pontificia Universidad Católica de Chile, Av. Vicuña Mackenna 4860, 7820436 Macul, Santiago, Chile*
[7] *Millenium Institute of Astrophysics, Av. Vicuña Mackenna 4860, 7820436 Macul, Santiago, Chile*
[8] *Max Planck Institute for Astronomy, Königstuhl 17, 69117 - Heidelberg, Germany*
[9] *Department of Physics, University of Rome Tor Vergata, Via della Ricerca Scientifica 1, I-00133 - Roma, Italy*
[10] *INAF - Astrophysical Observatory of Turin, Via Osservatorio 20, I-10025 - Pino Torinese, Italy*
[11] *Department of Physics, University of Texas at Dallas, Richardson, TX 75080, USA*
[12] *Astrochemistry Laboratory, Goddard Space Flight Center, NASA, 8800 Greenbelt Rd, Greenbelt, MD 20771, USA*
[13] *Harvard-Smithsonian Center for Astrophysics, 60 Garden St., Cambridge, MA 02138, USA*
[14] *Mississippi State University, Department of Physics & Astronomy, Hilbun Hall, Starkville, MS 39762, USA*
[15] *Department of Terrestrial Magnetism, Carnegie Institution for Science, Washington, DC 20015, USA*
[16] *The Observatories of the Carnegie Institution for Science, 813 Santa Barbara St, Pasadena, CA 91101, USA*
[17] *Astrophysics Research Centre, Queens University, Belfast, Northern Ireland, UK*
[18] *Perth Exoplanet Survey Telescope, Perth, Australia*
[19] *Australian Centre for Astrobiology, School of Physics, University of New South Wales, NSW 2052, Australia*
[20] *Exoplanetary Science at UNSW, School of Physics, University of New South Wales, NSW 2052, Australia*
[21] *Hungarian Astronomical Association, 1451 Budapest, Hungary*

## ABSTRACT

We report the discovery of ten transiting extrasolar planets by the HATSouth survey. The planets range in mass from the Super-Neptune HATS-62b, with $M_p < 0.179\,M_{\rm J}$, to the Super-Jupiter HATS-66b, with $M_p = 5.33\,M_{\rm J}$, and in size from the Saturn HATS-69b, with $R_p = 0.94\,R_{\rm J}$, to the inflated Jupiter HATS-67b, with $R_p = 1.69\,R_{\rm J}$. The planets have orbital periods between 1.6092 days (HATS-67b) and 7.8180 days (HATS-61b). The hosts are dwarf stars with masses ranging from $0.89\,M_\odot$ (HATS-69) to $1.56\,M_\odot$ (HATS-64), and have apparent magnitudes between $V = 12.276 \pm 0.020$ mag (HATS-68) and $V = 14.095 \pm 0.030$ mag (HATS-66). The Super-Neptune HATS-62b is the least massive planet discovered to date with a radius larger than Jupiter. Based largely on the Gaia DR2 distances and broad-band photometry, we identify three systems (HATS-62, -64, and -65) as having possible unresolved binary star companions. We discuss in detail our methods for incorporating the Gaia DR2 observations into our modeling of the system parameters, and into our blend analysis procedures.

Corresponding author: Joel Hartman
jhartman@astro.princeton.edu

* The HATSouth network is operated by a collaboration consisting of Princeton University (PU), the Max Planck Institute für Astronomie (MPIA), the Australian National University (ANU), and the Pontificia Universidad Católica de Chile (PUC). The station at Las Campanas Observatory (LCO) of the Carnegie Institute is operated by PU in conjunction with PUC, the station at the High Energy Spectroscopic Survey (H.E.S.S.) site is operated in conjunction with MPIA, and the station at Siding Spring Observatory (SSO) is operated jointly with ANU. Based in part on observations made with the MPG 2.2 m Telescope at the ESO Observatory in La Silla. Based on observations collected at the European Southern Observatory under ESO programmes 094.C-0428(A), 095.C-0367(A), 097.C-0571(A), 098.C-0292(A), 099.C-0374(A), 0100.C-0406(A), 0100.C-0406(B). This paper includes data gathered with the 6.5 meter Magellan Telescopes at Las Campanas Observatory, Chile. Based in part on observations made with the Anglo-Australian Telescope operated by the Australian Astronomical Observatory.

*Keywords:* planetary systems — stars: individual ( HATS-60, GSC 5818-00762, HATS-61, GSC 6459-01232 HATS-62, GSC 6922-00272 HATS-63, GSC 6470-00868 HATS-64, GSC 6613-00109 HATS-65, GSC 6884-00069 HATS-66, GSC 7096-01251 HATS-67, GSC 8229-00121 HATS-68, GSC 8479-00369 HATS-69, GSC ) techniques: spectroscopic, photometric

† Packard Fellow

‡ Bernoulli Fellow

Gruber Fellow

## 1. INTRODUCTION

This paper is part of a series of papers presenting the discovery and characterization of transiting exoplanetary systems by the HATSouth survey (Bakos et al. 2013). HATSouth is a wide-field ground-based photometric survey for transiting planets. Here we present the discovery, confirmation and characterization of ten new transiting planet systems by HATSouth. We number these systems as HATS-60 through HATS-69. The motivation for this work, and our methodology, have been discussed extensively elsewhere (e.g., Penev et al. 2013). Other works in this series from the past year include Bayliss et al. (2018a), Bento et al. (2018), Brahm et al. (2018), Henning et al. (2018), and Sarkis et al. (2018). Other currently active wide-field ground-based transit surveys include the following projects: WASP (Pollacco et al. 2006; recent discoveries include Demangeon et al. 2018, Hodžić et al. 2018, Lendl et al. 2018, Barkaoui et al. 2018, and Temple et al. 2018); HATNet (Bakos et al. 2004; Zhou et al. 2017 is the most recent published planet discovery); KELT (Pepper et al. 2007; recent discoveries include Siverd et al. 2018, Johnson et al. 2018, and Labadie-Bartz et al. 2018); The Qatar Exoplanet Survey (Alsubai et al. 2013; Alsubai et al. 2018 is a discovery from the past year); NGTS (Wheatley et al. 2018; recent discoveries include Bayliss et al. 2018b, Raynard et al. 2018, and Günther et al. 2018); and MASCARA (Talens et al. 2017; Talens et al. 2018 is a discovery from the past year). Dedicated space missions to find transiting planets include *Kepler* (Borucki et al. 2010), *K2* (Howell et al. 2014), *CoRoT* (Auvergne et al. 2009) and the recently launched *TESS* mission (Ricker et al. 2015). The planets presented here contribute to our growing understanding of planetary systems in the Galaxy.

In this work we take advantage of the recent release of high-precision geometric parallax measurements for all of these objects by the *Gaia* mission (Gaia Collaboration et al. 2016, 2018). These distance measurements enable a much more precise characterization of the systems than has heretofore been possible for most such objects. The distances also allow us to confirm planetary systems for which we had previously been unable to unambiguously rule out the possibility of their being blended stellar eclipsing binary systems, and to detect possible unresolved binary star companions to the planetary host stars.

## 2. OBSERVATIONS

Figures 1 through 10 show the observations collected for HATS-60 through HATS-69, respectively. Each figure shows the HATSouth light curve used to detect the transits, the ground-based follow-up transit light curves, the high-precision RVs and spectral line bisector spans (BSs), and the catalog broad-band photometry, including parallax corrections from Gaia DR2, used in characterizing the host stars. Below we describe the observations of these objects that were collected by our team.

### 2.1. *Photometric detection*

All ten systems presented here were initially detected as transiting planet candidates based on observations by the HATSouth network. The operations of the network are described in Bakos et al. (2013), while our methods for reducing the data to trend-filtered light curves (filtered using the method of Kovács et al. 2005) and identifying transiting planet signals (using the Box-fitting Least Squares or BLS method; Kovács et al. 2002) are described in Penev et al. (2013). The HATSouth observations of each system are summarized in Table 1, while the light curve data are made available in Table 5.

We also searched the light curves for other periodic signals using the Generalized Lomb-Scargle method (GLS; Zechmeister & Kürster 2009), and for additional transit signals by applying a second iteration of BLS. Both of these searches were performed on the residual light curves after subtracting the best-fit primary transit models.

Table 2 gives the GLS results for each target, including the peak period, false alarm probability, semi-amplitude, and 95% confidence upper bound on the semi-amplitude of the highest significance periodic signal in the light curves. Here the false alarm probabilities are calculated by performing bootstrap simulations. HATS-61 shows evidence for a $P = 28.54$ day periodic signal with a semi-amplitude of 0.32 mmag. The false alarm probability of this detection is $10^{-3.7}$. This may correspond to the photometric rotation period of this $5630 \pm 71$ K star. The star has $v \sin i = 3.52 \pm 0.42\,\mathrm{km\,s^{-1}}$, which gives an upper limit of $24.0 \pm 3.2$ days on the equatorial rotation period. The photometric period of 28.54 days is above the limit at the $\sim 1.4\sigma$ level, so would be consistent with $v \sin i$ if it has been slightly overestimated and the planet orbital axis is aligned with the stellar rotation axis, or if there is modest differential rotation and the spots are at a more slowly rotating latitude on the star. None of the other targets shows a statistically significant sinusoidal periodic signal.

Table 3 gives the BLS results for additional transit signals that may be present in the HATSouth light curve of each target, including the period, transit depth, transit duration, and S/N for the top peak in the BLS spectrum. HATS-62 shows a possible transit signal with a period of 12.9395 days, duration of 0.339 days and a depth of 5.5 mmag. The S/N is a modest 7.5, and the signal is most likely a false alarm. Observations of this system already carried out by the NASA TESS mission will confirm or refute it. The reference mid transit time is $T_C = 2455099.556$ BJD. None of the other objects show evidence for additional transit signals in their HATSouth light curves.

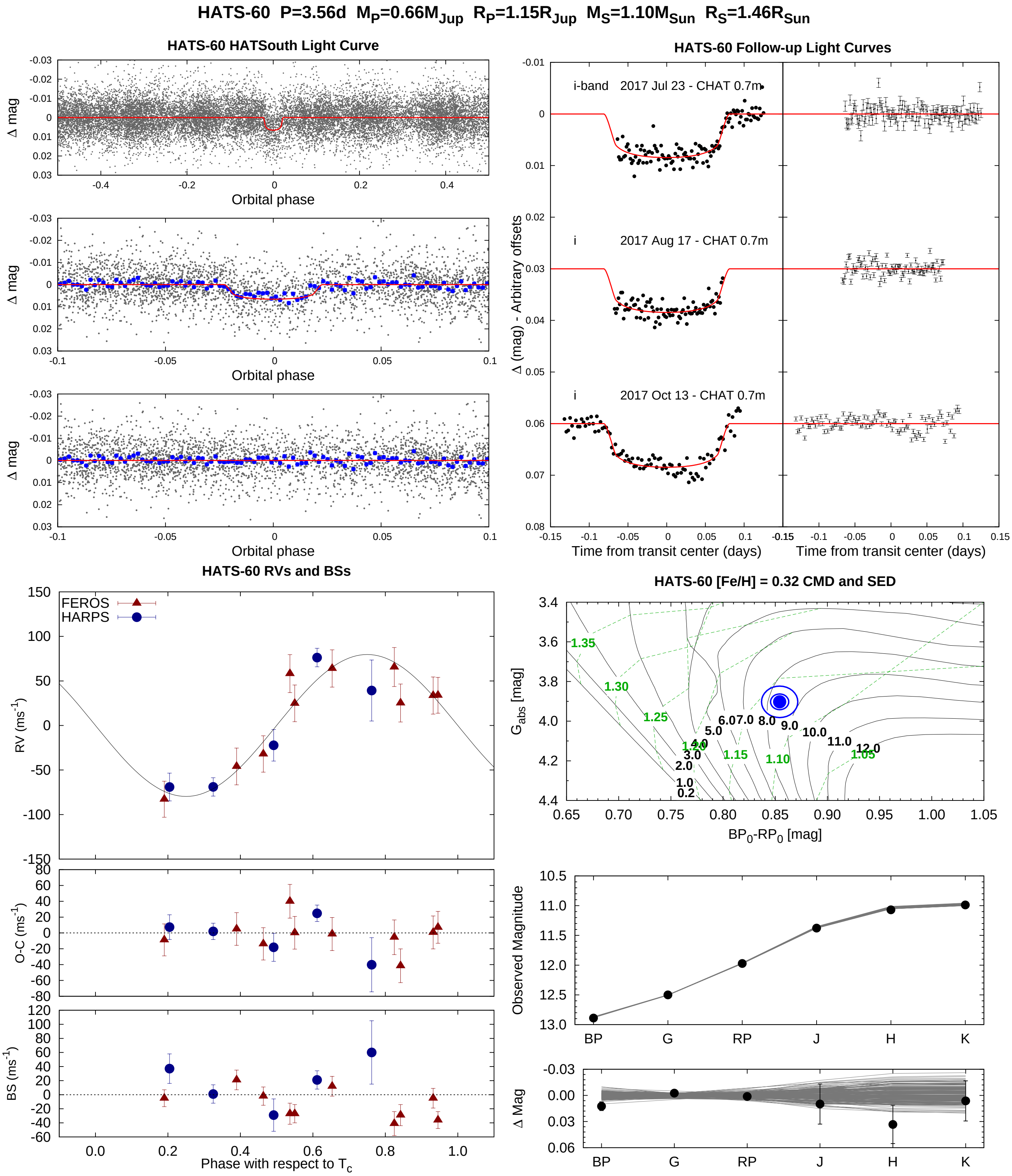

**Figure 1.** Observations used to confirm the transiting planet system HATS-60. *Top Left:* Phase-folded unbinned HATSouth light curve. The top panel shows the full light curve, the middle panel shows the light curve zoomed-in on the transit, and the bottom panel shows the residuals from the best-fit model zoomed-in on the transit. The solid lines show the model fits to the light curves. The dark filled circles show the light curves binned in phase with a bin size of 0.002. The slight systematic discrepancy between the model and binned values apparent in the middle panel is an artifact of plotting data from multiple HATSouth fields with differing effective transit dilution factors. The quality of the fit in this case is best judged by inspection of the residuals shown in the bottom panel. (Caption continued on next page).

**Figure 1.** (Caption continued from previous page) *Top Right:* Unbinned follow-up transit light curves corrected for instrumental trends fitted simultaneously with the transit model, which is overplotted. The dates, filters and instruments used are indicated. The residuals are shown on the right-hand-side in the same order as the original light curves. The error bars represent the photon and background shot noise, plus the readout noise. Note that these uncertainties are scaled up in the fitting procedure to achieve a reduced $\chi^2$ of unity, but the uncertainties shown in the plot have not been scaled. *Bottom Left:* High-precision RVs phased with respect to the mid-transit-time. The instruments used are labelled in the plot. The top panel shows the phased measurements together with the best-fit model. The center-of-mass velocity has been subtracted. The second panel shows the velocity $O-C$ residuals. The error bars include the estimated jitter. The third panel shows the bisector spans. *Bottom Right:* Color-magnitude diagram (CMD) and spectral energy distribution (SED). The top panel shows the absolute $G$ magnitude vs. the de-reddened $BP - RP$ color compared to theoretical isochrones (black lines) and stellar evolution tracks (green lines) from the PARSEC models interpolated at the spectroscopically determined metallicity of the host. The age of each isochrone is listed in black in Gyr, while the mass of each evolution track is listed in green in solar mass units. The filled blue circles show the measured reddening- and distance-corrected values from Gaia DR2, while the blue lines indicate the $1\sigma$ and $2\sigma$ confidence regions, including the estimated systematic errors in the photometry. The middle panel shows the SED as measured via broadband photometry through the six listed filters. Here we plot the observed magnitudes without correcting for distance or extinction. Overplotted are 200 model SEDs randomly selected from the MCMC posterior distribution produced through the global analysis. The model makes use of the predicted absolute magnitudes in each bandpass from the PARSEC isochrones, the distance to the system (constrained largely via Gaia DR2) and extinction (constrained largely via the MWDUST 3D Galactic extinction model). The bottom panel shows the $O-C$ residuals from the best-fit model SED.

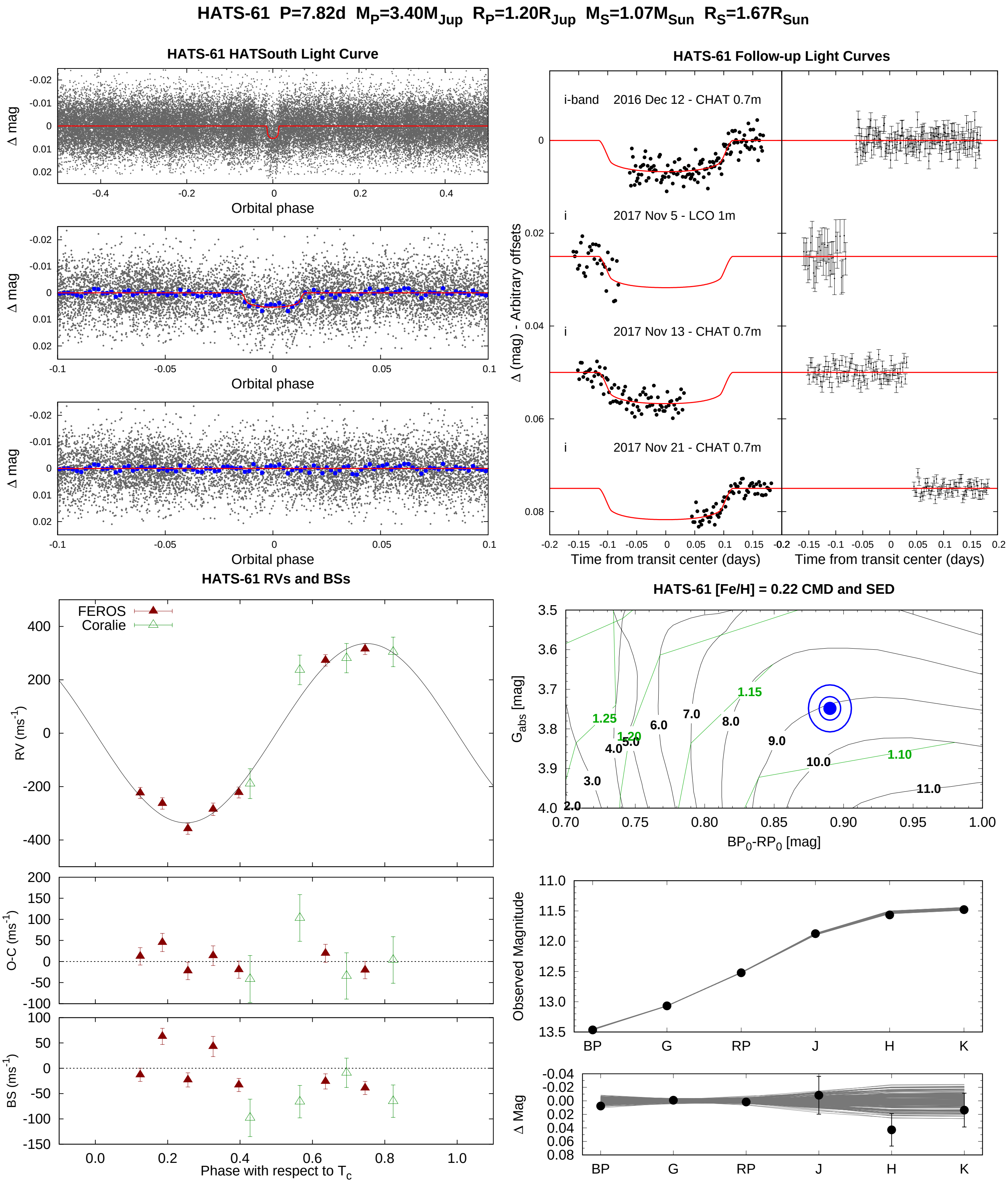

**Figure 2.** Same as Figure 1, here we show the observations of HATS-61.

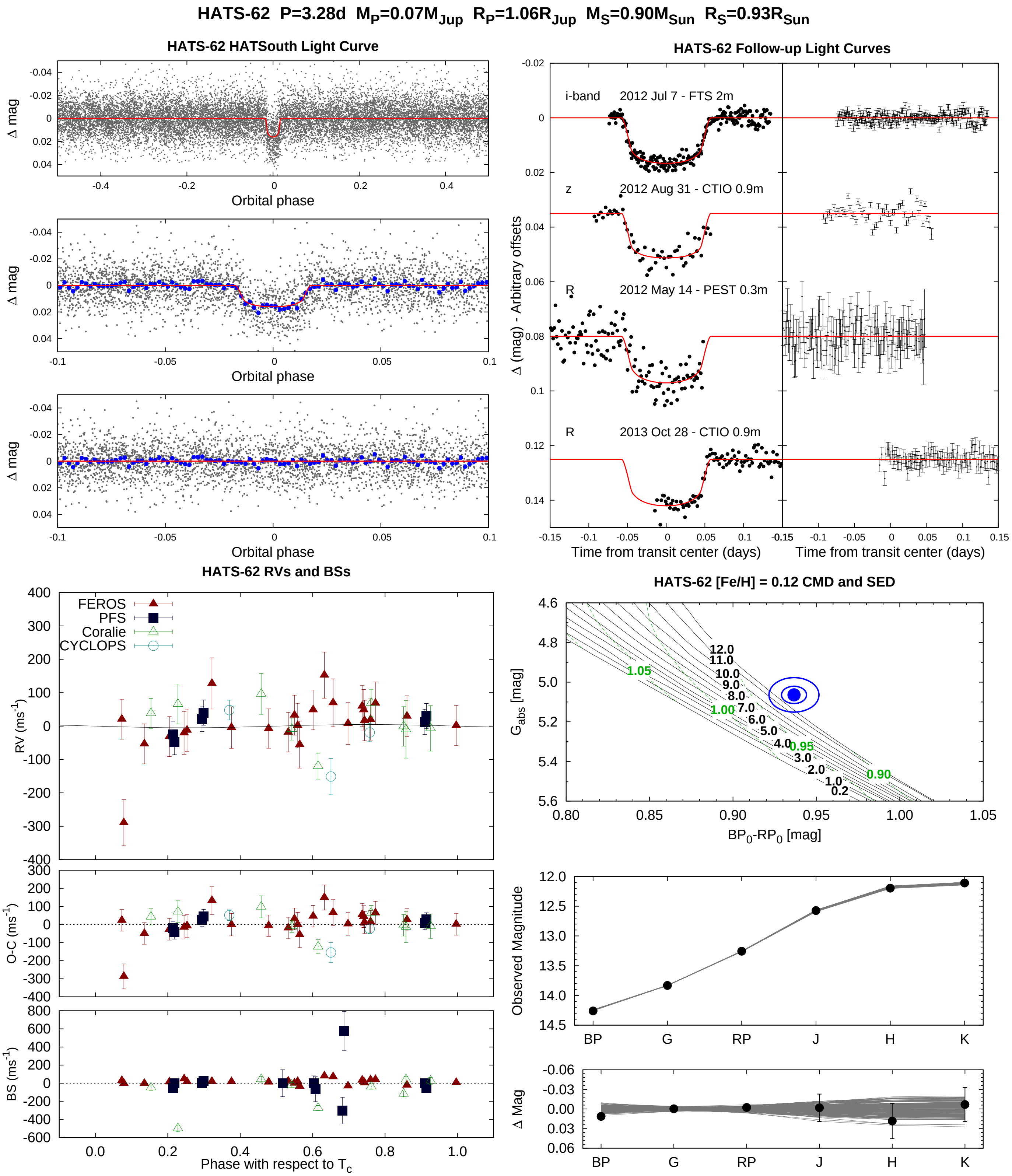

**Figure 3.** Same as Figure 1, here we show the observations of HATS-62. Note that for some observations accurate bisector spans could not be measured, but RVs could be measured. For the $I_2$-free PFS observations we measured bisector spans, but not RVs.

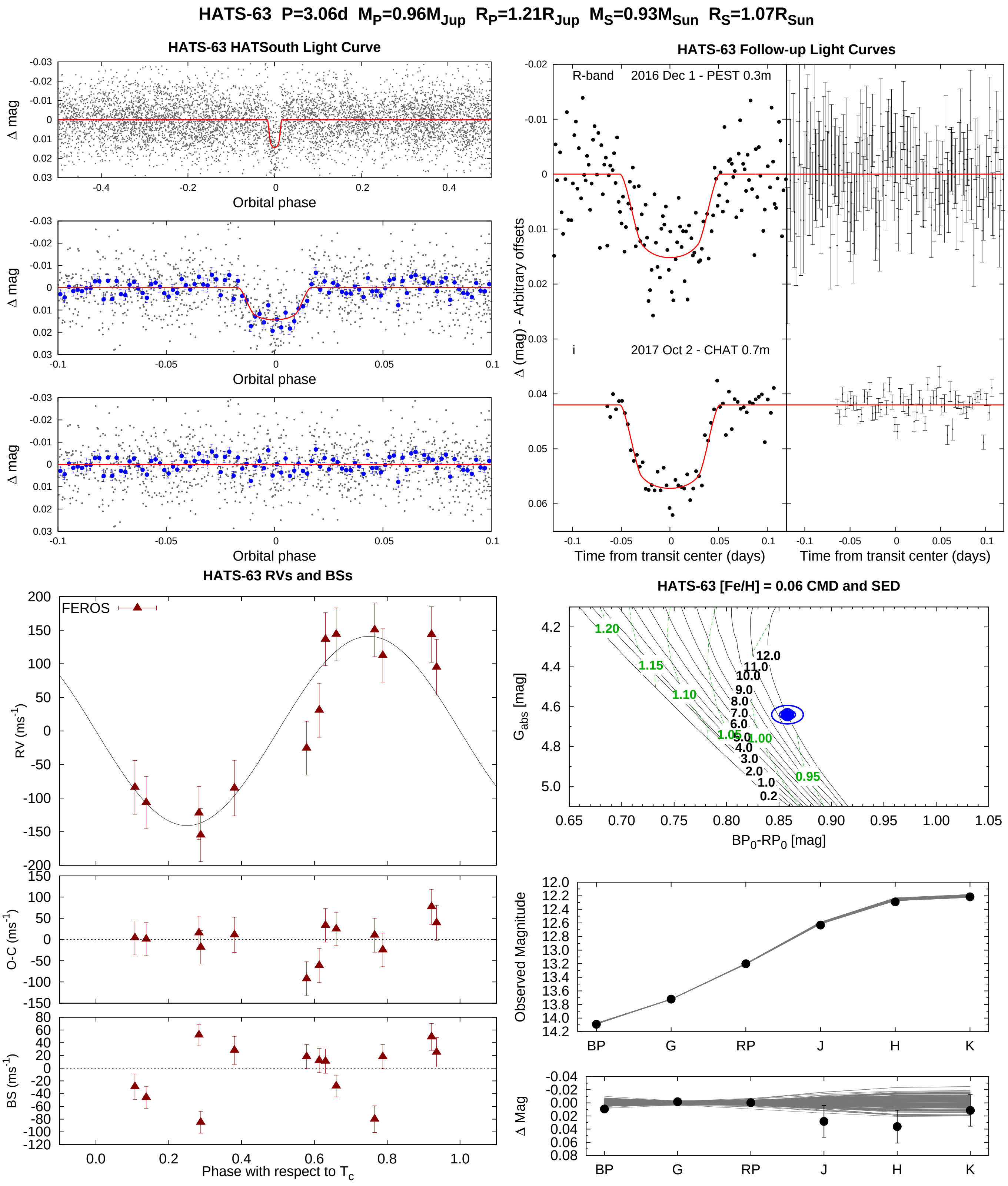

**Figure 4.** Same as Figure 1, here we show the observations of HATS-63.

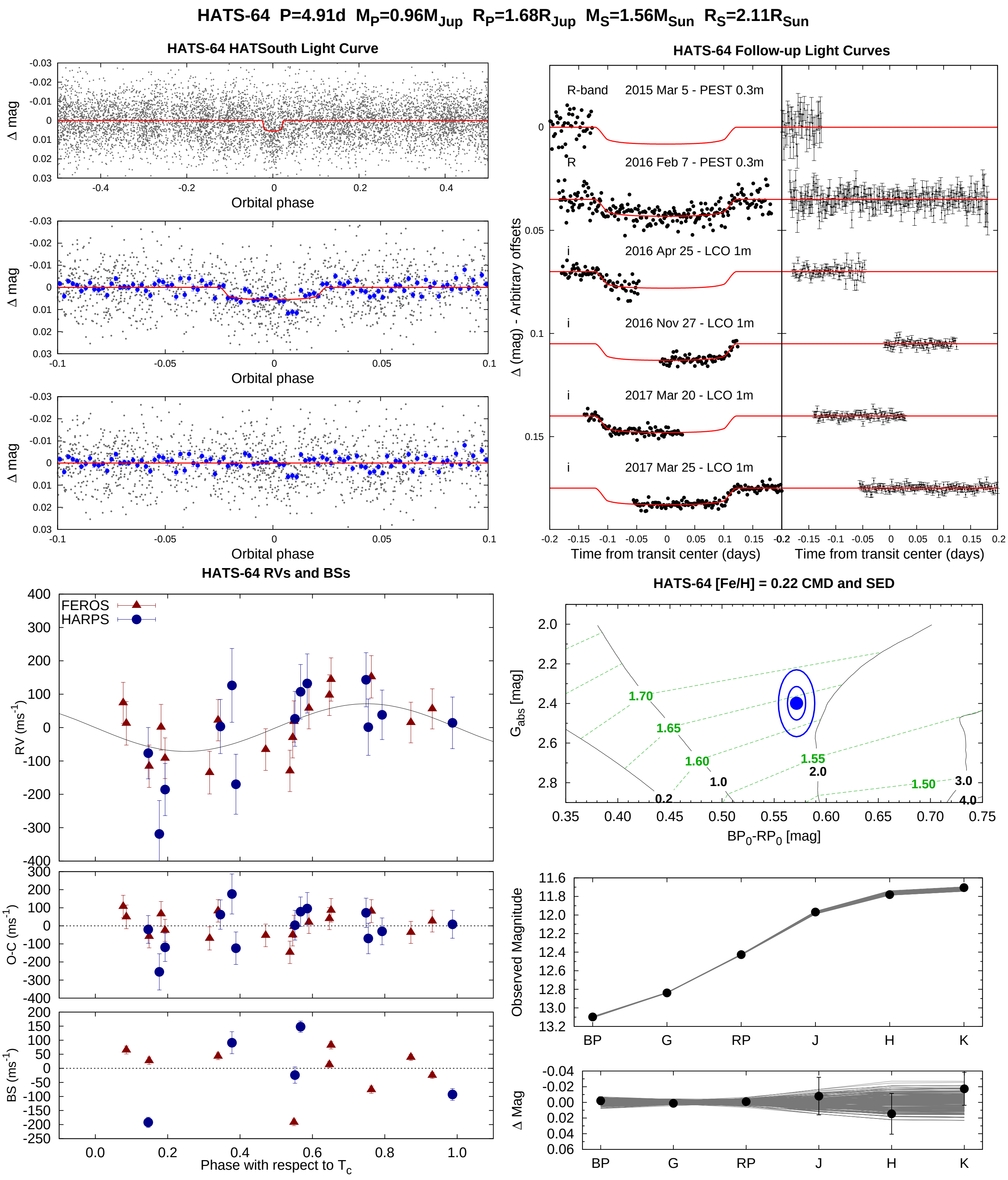

**Figure 5.** Same as Figure 1, here we show the observations of HATS-64. Note that for some observations accurate bisector spans could not be measured, but RVs could be measured.

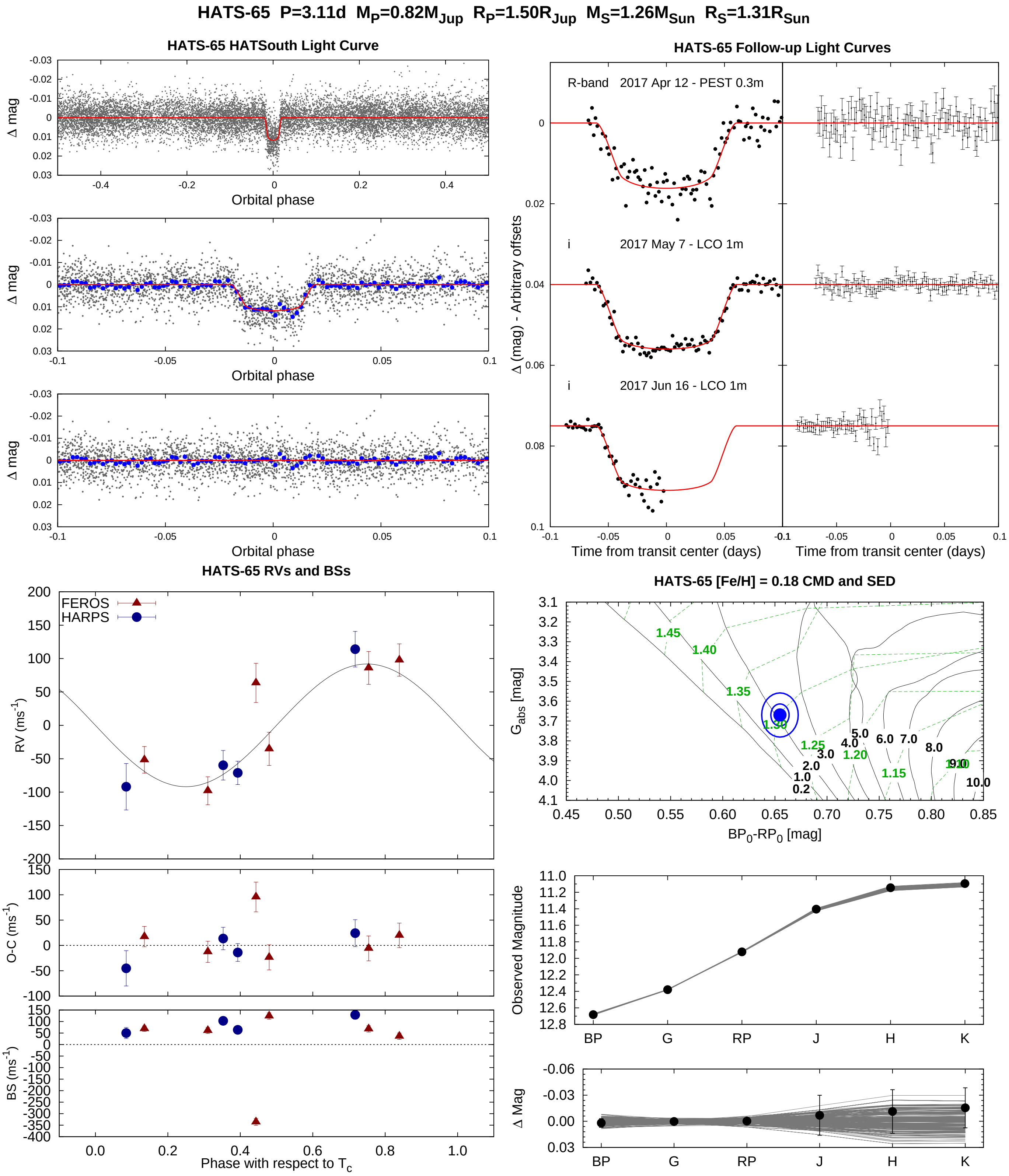

**Figure 6.** Same as Figure 1, here we show the observations of HATS-65.

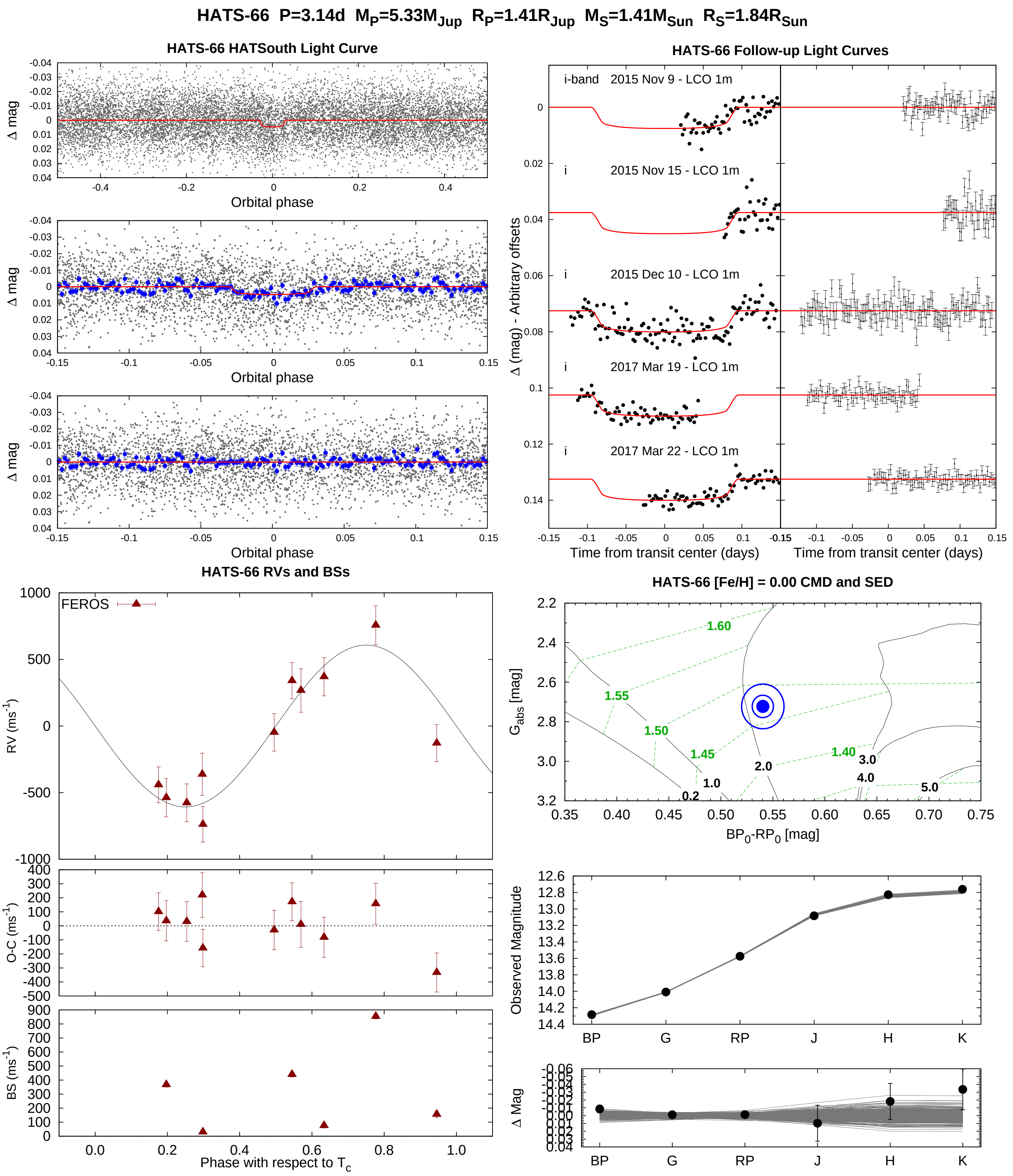

**Figure 7.** Same as Figure 1, here we show the observations of HATS-66. Note that for some observations accurate bisector spans could not be measured, but RVs could be measured.

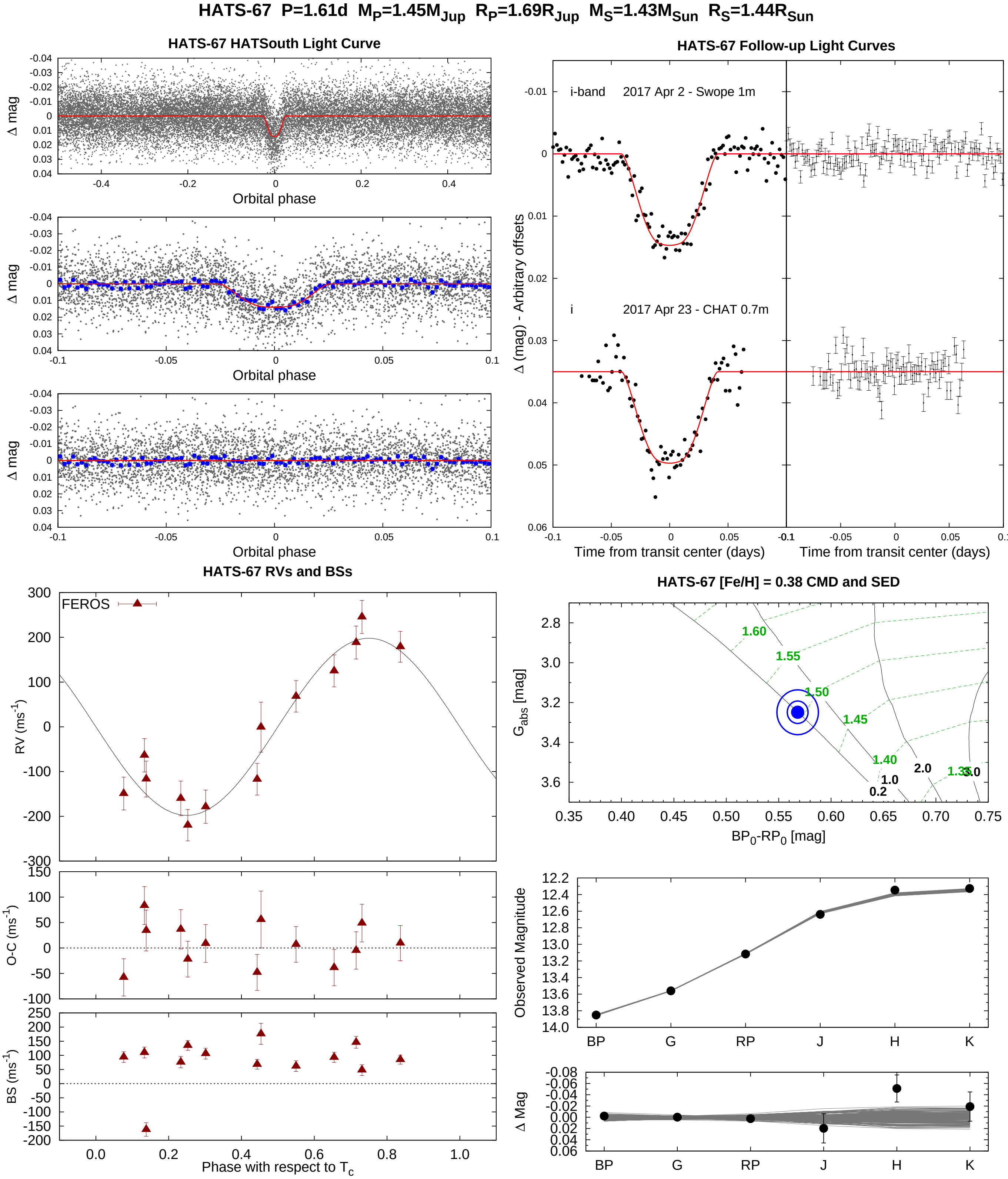

**Figure 8.** Same as Figure 1, here we show the observations of HATS-67.

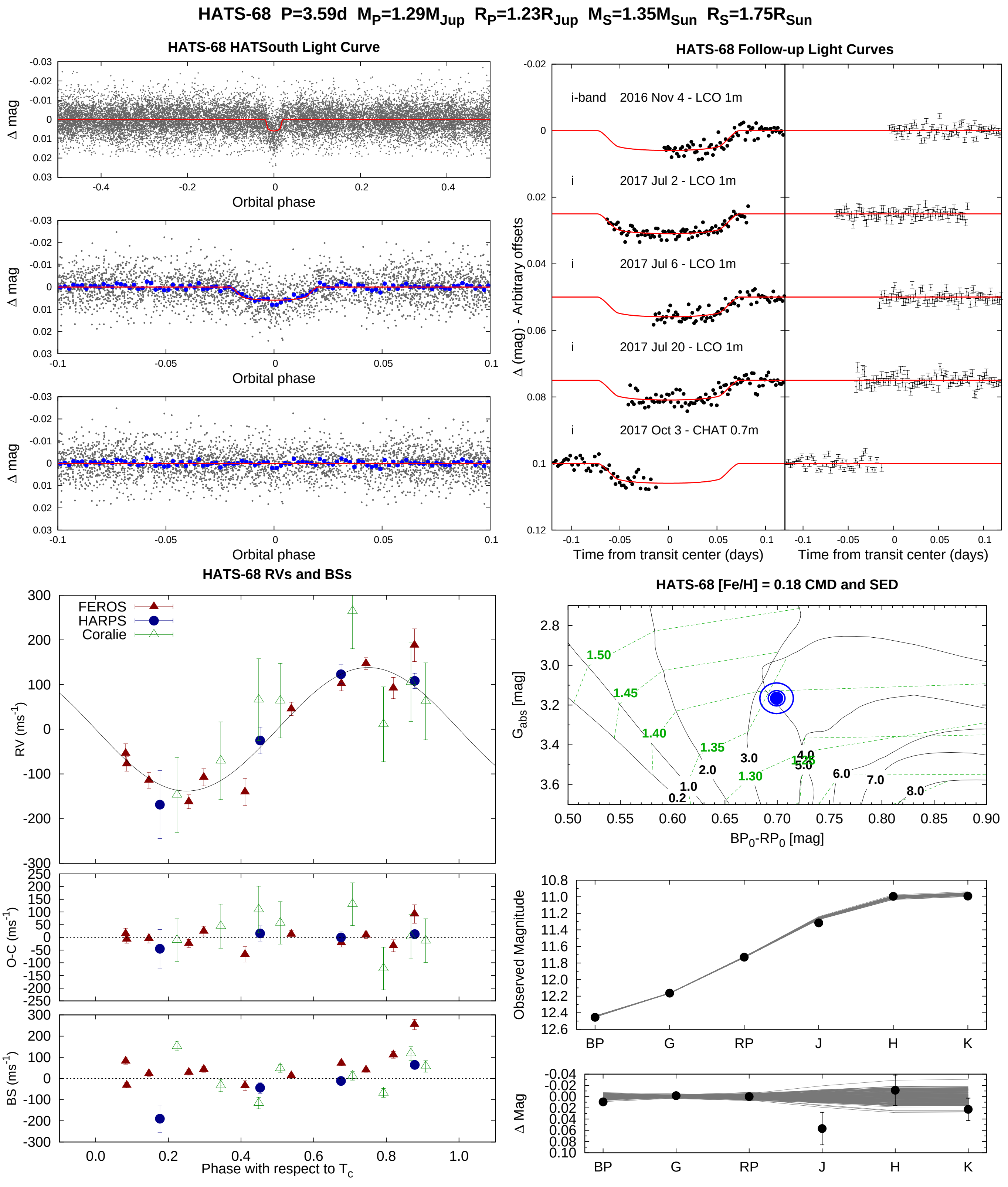

**Figure 9.** Same as Figure 1, here we show the observations of HATS-68.

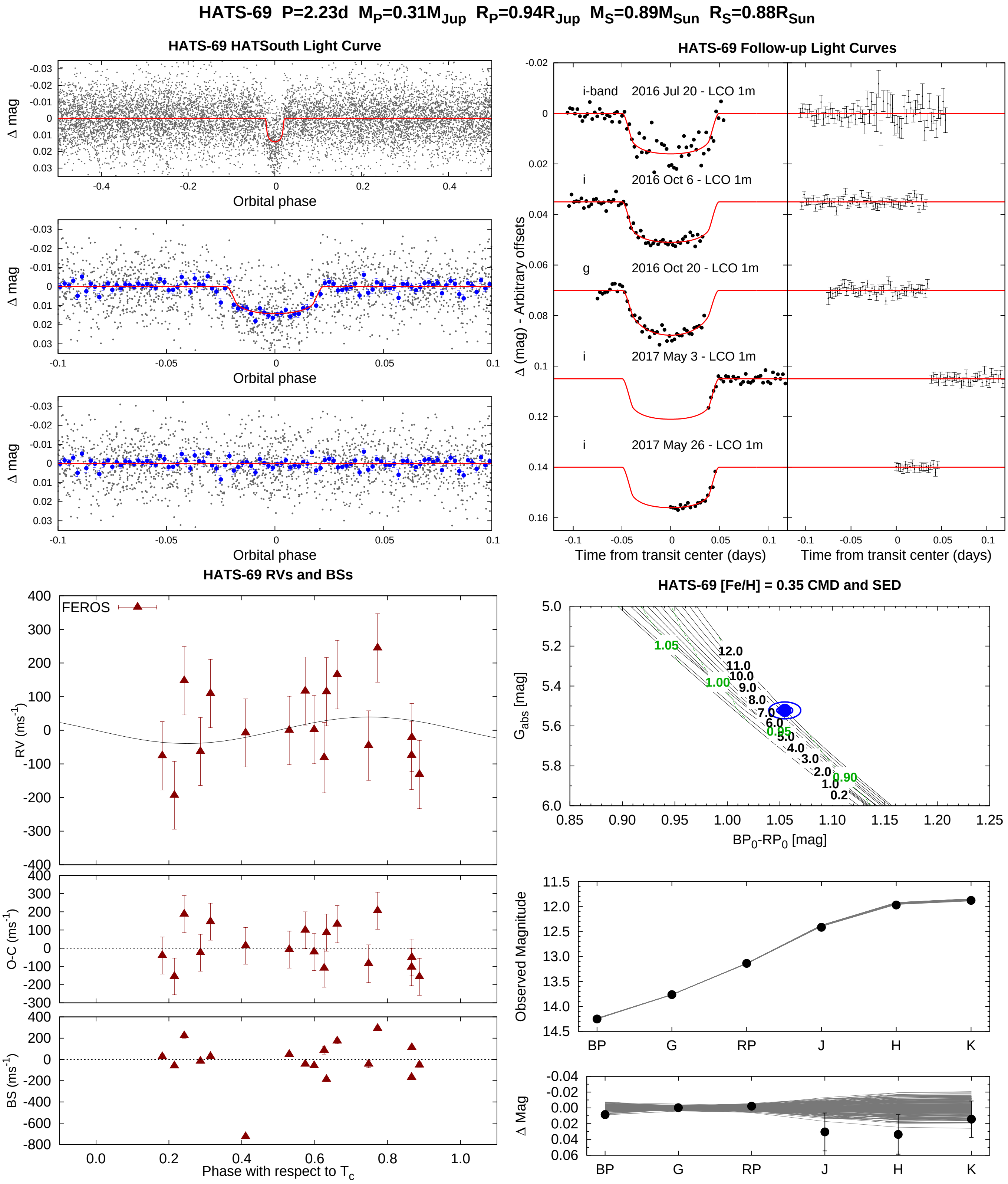

**Figure 10.** Same as Figure 1, here we show the observations of HATS-69.

**Table 1**. Summary of photometric observations

| Instrument/Field[a] | Date(s) | # Images | Cadence[b] (sec) | Filter | Precision[c] (mmag) |
|---|---|---|---|---|---|
| **HATS-60** | | | | | |
| HS-1/G537.3 | 2016 Nov–2016 Dec | 292 | 350 | $r$ | 9.2 |
| HS-3/G537.3 | 2016 Jun–2016 Dec | 5597 | 324 | $r$ | 6.1 |
| HS-5/G537.3 | 2016 Jun–2016 Dec | 3216 | 365 | $r$ | 6.8 |
| HS-1/G537.4 | 2016 Jun–2016 Dec | 4101 | 333 | $r$ | 9.7 |
| HS-3/G537.4 | 2016 Oct–2016 Dec | 28 | 1179 | $r$ | 8.2 |
| HS-5/G537.4 | 2016 Jun–2016 Dec | 3334 | 365 | $r$ | 8.4 |
| CHAT 0.7 m | 2017 Jul 23 | 115 | 142 | $i$ | 1.5 |
| CHAT 0.7 m | 2017 Aug 17 | 85 | 143 | $i$ | 1.4 |
| CHAT 0.7 m | 2017 Oct 13 | 93 | 210 | $i$ | 1.4 |
| **HATS-61** | | | | | |
| HS-1/G548.4 | 2014 Sep–2015 Apr | 6601 | 287 | $r$ | 7.1 |
| HS-2/G548.4 | 2014 Jun–2015 Apr | 7650 | 348 | $r$ | 6.8 |
| HS-3/G548.4 | 2014 Sep–2015 Mar | 5313 | 352 | $r$ | 6.7 |
| HS-4/G548.4 | 2014 Jun–2015 Mar | 6013 | 352 | $r$ | 6.4 |
| HS-5/G548.4 | 2014 Sep–2015 Mar | 5007 | 359 | $r$ | 7.2 |
| HS-6/G548.4 | 2014 Jul–2015 Mar | 6002 | 351 | $r$ | 6.9 |
| CHAT 0.7 m | 2016 Dec 12 | 128 | 146 | $i$ | 1.9 |
| LCO 1 m/MCD/sinistro | 2017 Nov 05 | 30 | 224 | $i$ | 2.8 |
| CHAT 0.7 m | 2017 Nov 13 | 79 | 203 | $i$ | 1.7 |
| CHAT 0.7 m | 2017 Nov 21 | 60 | 200 | $i$ | 1.2 |
| **HATS-62** | | | | | |
| HS-2/G582.1 | 2009 Sep–2010 Sep | 5649 | 284 | $r$ | 12.6 |
| HS-4/G582.1 | 2009 Sep–2010 Sep | 8925 | 288 | $r$ | 12.3 |
| HS-6/G582.1 | 2010 Aug–2010 Sep | 201 | 290 | $r$ | 11.5 |
| FTS 2 m | 2012 Jul 07 | 225 | 80 | $i$ | 1.8 |
| CTIO 0.9 m | 2012 Aug 31 | 54 | 240 | $z$ | 3.2 |
| PEST 0.3 m | 2013 May 14 | 141 | 130 | $R_C$ | 5.3 |
| CTIO 0.9 m | 2013 Oct 28 | 91 | 177 | $R$ | 2.4 |
| **HATS-63** | | | | | |
| HS-1/G597.2 | 2013 Sep–2014 Mar | 1555 | 286 | $r$ | 10.0 |
| HS-3/G597.2 | 2013 Sep–2014 Feb | 4487 | 285 | $r$ | 10.4 |
| PEST 0.3 m | 2016 Dec 01 | 151 | 132 | $R_C$ | 6.1 |
| CHAT 0.7 m | 2017 Oct 02 | 57 | 267 | $i$ | 2.3 |
| **HATS-64** | | | | | |
| HS-2/G606.3 | 2012 Feb–2012 Jun | 3132 | 291 | $r$ | 8.8 |
| HS-4/G606.3 | 2012 Feb–2012 Jun | 2750 | 300 | $r$ | 9.9 |
| HS-6/G606.3 | 2012 Feb–2012 Jun | 1143 | 299 | $r$ | 10.1 |
| DK 1.54 m | 2014 Mar 16 | 229 | 144 | $R$ | 1.4 |
| PEST 0.3 m | 2015 Mar 05 | 202 | 132 | $R_C$ | 5.2 |
| PEST 0.3 m | 2016 Feb 07 | 224 | 132 | $R_C$ | 3.9 |
| LCO 1 m/CTIO/sinistro | 2016 Apr 25 | 70 | 159 | $i$ | 2.5 |
| LCO 1 m/CTIO/sinistro | 2016 Nov 27 | 73 | 160 | $i$ | 1.5 |

*Table 1 continued*

**Table 1** *(continued)*

| Instrument/Field[a] | Date(s) | # Images | Cadence[b] (sec) | Filter | Precision[c] (mmag) |
|---|---|---|---|---|---|
| LCO 1 m/CTIO/sinistro | 2017 Mar 20 | 91 | 160 | $i$ | 1.4 |
| LCO 1 m/CTIO/sinistro | 2017 Mar 25 | 140 | 160 | $i$ | 1.4 |
| **HATS-65** | | | | | |
| HS-1/G625.2 | 2012 Jun–2012 Oct | 4694 | 291 | $r$ | 6.0 |
| HS-3/G625.2 | 2012 Jun–2012 Oct | 5359 | 293 | $r$ | 5.6 |
| HS-5/G625.2 | 2012 Jun–2012 Oct | 1752 | 293 | $r$ | 6.4 |
| PEST 0.3 m | 2017 Apr 12 | 91 | 132 | $R_C$ | 3.0 |
| LCO 1 m/SSO/sinistro | 2017 May 07 | 96 | 161 | $i$ | 1.3 |
| LCO 1 m/SAAO/sinistro | 2017 Jun 16 | 46 | 161 | $i$ | 1.8 |
| **HATS-66** | | | | | |
| HS-1/G601.1 | 2011 Aug–2012 Jan | 4779 | 296 | $r$ | 13.6 |
| HS-3/G601.1 | 2011 Aug–2012 Jan | 4081 | 296 | $r$ | 12.9 |
| HS-5/G601.1 | 2011 Aug–2012 Jan | 3088 | 290 | $r$ | 12.4 |
| LCO 1 m/SBIG | 2015 Nov 09 | 90 | 192 | $i$ | 2.6 |
| LCO 1 m/SBIG | 2015 Nov 15 | 38 | 193 | $i$ | 4.1 |
| LCO 1 m/SBIG | 2015 Dec 10 | 118 | 193 | $i$ | 3.5 |
| LCO 1 m/SAAO/sinistro | 2017 Mar 19 | 61 | 221 | $i$ | 2.0 |
| LCO 1 m/CTIO/sinistro | 2017 Mar 22 | 69 | 220 | $i$ | 1.8 |
| **HATS-67** | | | | | |
| HS-4/G698.1 | 2015 May–2015 Jul | 5 | 499 | $r$ | 12.1 |
| HS-6/G698.1 | 2015 Dec–2016 Jun | 4431 | 344 | $r$ | 12.1 |
| HS-2/G698.4 | 2015 Mar–2016 May | 2482 | 352 | $r$ | 11.4 |
| HS-4/G698.4 | 2015 Mar–2016 Jun | 6894 | 324 | $r$ | 11.0 |
| HS-6/G698.4 | 2015 Mar–2016 Jun | 5759 | 343 | $r$ | 10.6 |
| Swope 1 m | 2017 Apr 02 | 139 | 140 | $i$ | 1.7 |
| CHAT 0.7 m | 2017 Apr 23 | 77 | 149 | $i$ | 2.2 |
| **HATS-68** | | | | | |
| HS-1/G755.3 | 2011 Jul–2012 Oct | 5119 | 292 | $r$ | 6.8 |
| HS-3/G755.3 | 2011 Jul–2012 Oct | 4896 | 287 | $r$ | 6.2 |
| HS-5/G755.3 | 2011 Jul–2012 Oct | 5875 | 296 | $r$ | 5.8 |
| LCO 1 m/SAAO/sinistro | 2016 Nov 04 | 67 | 160 | $i$ | 1.5 |
| LCO 1 m/SAAO/sinistro | 2017 Jul 02 | 79 | 161 | $i$ | 1.2 |
| LCO 1 m/SSO/sinistro | 2017 Jul 06 | 77 | 164 | $i$ | 1.4 |
| LCO 1 m/SAAO/sinistro | 2017 Jul 20 | 94 | 164 | $i$ | 1.7 |
| CHAT 0.7 m | 2017 Oct 03 | 71 | 184 | $i$ | 1.6 |
| **HATS-69** | | | | | |
| HS-2/G778.4 | 2011 May–2012 Nov | 3052 | 287 | $r$ | 12.3 |
| HS-4/G778.4 | 2011 Jul–2012 Nov | 3686 | 298 | $r$ | 11.6 |
| HS-6/G778.4 | 2011 Apr–2012 Oct | 2325 | 298 | $r$ | 11.2 |
| LCO 1 m/CTIO/sinistro | 2016 Jul 20 | 63 | 219 | $i$ | 3.7 |
| LCO 1 m/CTIO/sinistro | 2016 Oct 06 | 55 | 219 | $i$ | 1.5 |
| LCO 1 m/SBIG | 2016 Oct 20 | 44 | 220 | $g$ | 1.7 |
| LCO 1 m/CTIO/sinistro | 2017 May 03 | 33 | 220 | $i$ | 1.4 |
| LCO 1 m/SSO/sinistro | 2017 May 26 | 18 | 221 | $i$ | 0.9 |

*Table 1 continued*

**Table 1** *(continued)*

| Instrument/Field[a] | Date(s) | # Images | Cadence[b] (sec) | Filter | Precision[c] (mmag) |
|---|---|---|---|---|---|

[a] For HATSouth data we list the HATSouth unit, CCD and field name from which the observations are taken. HS-1 and -2 are located at Las Campanas Observatory in Chile, HS-3 and -4 are located at the H.E.S.S. site in Namibia, and HS-5 and -6 are located at Siding Spring Observatory in Australia. Each unit has 4 ccds. Each field corresponds to one of 838 fixed pointings used to cover the full $4\pi$ celestial sphere. All data from a given HATSouth field and CCD number are reduced together, while detrending through External Parameter Decorrelation (EPD) is done independently for each unique unit+CCD+field combination.

[b] The median time between consecutive images rounded to the nearest second. Due to factors such as weather, the day–night cycle, guiding and focus corrections the cadence is only approximately uniform over short timescales.

[c] The RMS of the residuals from the best-fit model.

**Table 2**. GLS Search for Periodic Signals in HATSouth Light Curves

| System | Peak Period (days) | $\log_{10}$(FAP) | Amplitude (mmag) | Amplitude 95% Upper Limit (mmag) |
|---|---|---|---|---|
| HATS-60 | 0.46558049 | -0.35 | 0.43 | 0.57 |
| HATS-61 | 28.53996289 | -3.70 | 0.32 | 0.43 |
| HATS-62 | 0.01724177 | -1.03 | 0.78 | 1.1 |
| HATS-63 | 0.14945790 | -0.25 | 1.1 | 1.6 |
| HATS-64 | 0.07413442 | -0.43 | 0.92 | 1.2 |
| HATS-65 | 0.01288701 | -0.33 | 0.42 | 0.65 |
| HATS-66 | 0.01274483 | -0.02 | 0.94 | 1.4 |
| HATS-67 | 8.85543462 | -0.61 | 0.63 | 0.94 |
| HATS-68 | 0.99279159 | -0.79 | 0.43 | 0.61 |
| HATS-69 | 0.06501927 | -0.57 | 0.86 | 1.1 |

**Table 3**. BLS Search for Additional Transit Signals in HATSouth Light Curves

| System | Peak Period (days) | Transit Depth (mmag) | Transit Duration (days) | S/N |
|---|---|---|---|---|
| HATS-60 | 1.61123533 | 2.5 | 0.0506 | 6.5 |
| HATS-61 | 88.89871719 | 0.74 | 10.3 | 5.7 |
| HATS-62 | 12.93945856 | 5.5 | 0.339 | 7.5 |
| HATS-63 | 5.26120155 | 4.7 | 0.241 | 6.8 |
| HATS-64 | 0.31911273 | 4.8 | 0.00670 | 6.2 |
| HATS-65 | 0.41484153 | 1.1 | 0.0437 | 5.3 |
| HATS-66 | 0.22055123 | 3.9 | 0.00772 | 6.0 |
| HATS-67 | 18.14714871 | 1.6 | 2.07 | 5.6 |
| HATS-68 | 2.97022674 | 1.4 | 0.171 | 5.9 |
| HATS-69 | 0.11051985 | 4.2 | 0.00309 | 5.6 |

### 2.2. *Spectroscopic Observations*

The spectroscopic observations carried out to confirm and characterize each of the transiting planet systems are summarized in Table 4. The facilities used include FEROS on the MPG 2.2 m (all 10 targets; 138 observations total; Kaufer & Pasquini 1998), Coralie on the Euler 1.2 m (5 targets; 28 observations total; Queloz et al. 2001), HARPS on the ESO 3.6 m (4 targets; 27 observations total; Mayor et al. 2003), WiFeS on the ANU 2.3 m (5 targets; 18 observations total; Dopita et al. 2007),

PFS on the Magellan 6.5 m (1 target; 10 observations; Crane et al. 2010), UVES on the VLT UT2 8 m (3 targets; 3 observations; Dekker et al. 2000), and CYCLOPS on the AAT 3.9 m (1 target; 3 observations; Horton et al. 2012).

The FEROS, Coralie, HARPS and UVES observations were reduced to wavelength-calibrated spectra and high-precision RV and Bisector Span (BS) measurements using the CERES pipeline (Brahm et al. 2017a). We note that the RV and BS uncertainties do not include potential systematic errors due to sky contamination, which are particularly large for the faint, rapidly rotating star HATS-66. We also used the FEROS and UVES observations to determine high-precision stellar atmospheric parameters, including the effective temperture $T_{\rm eff\star}$, surface gravity $\log g$, metallicity [Fe/H], and $v \sin i$ via the ZASPE package (Brahm et al. 2017b). The UVES observations were used for this purpose for HATS-62, HATS-63 and HATS-66, while the FEROS observations were used for this purpose for the other seven systems. The UVES observations were obtained solely for measuring these atmospheric parameters, and were not included in the RV analysis of each system.

The WiFeS observations, which were used for reconnaissance of the targets, were reduced following Bayliss et al. (2013). For each target observed, we obtained a single spectrum at resolution $R \equiv \Delta\lambda/\lambda \approx 3000$ from which we estimated the effective temperature, $\log g$ and [Fe/H] of the star. Two to four observations at $R \approx 7000$ were also obtained to search for any large amplitude radial velocity variations at the $\sim 4\,{\rm km\,s^{-1}}$ level, which would indicate a stellar mass companion.

The PFS observations of HATS-62 include eight observations through an $I_2$ cell, and two observations without the cell used to construct a spectral template. The observations were reduced to spectra and used to determine high precision relative RV measurements following Butler et al. (1996). Spectral line bisector spans and their uncertainties were measured as described by Jordán et al. (2014) and Brahm et al. (2017a).

The CYCLOPS observations of HATS-62 were reduced to spectra and RV measurements following Addison et al. (2013)

The high-precision RV and BS measurements are given in Table 19 for all ten systems at the end of the paper.

### 2.3. *Photometric follow-up observations*

Follow-up higher-precision ground-based photometric transits observations were obtained for all ten systems, as summarized in Table 1. The facilities used for this purpose include: the Chilean-Hungarian Automated Telescope (CHAT) 0.7 m telescope at Las Campanas Observatory, Chile (6 transits of 4 targets; Jordán et al. 2018); 1 m telescopes from the Las Cumbres Observatory (LCO) network, including units at McDonald Observatory (MCD) in Texas, at Cerro Telolo Inter-American Observatory (CTIO) in Chile, at Siding Spring Observatory (SSO) in Australia, and at the South African Astronomical Observatory (SAAO) in South Africa (21 transits of 6 targets altogether; Brown et al. 2013); the 2 m Faulkes Telescope South (FTS) operated at SSO by LCO (one transit of one target); the SMARTS CTIO 0.9 m telescope (2 transits of 1 target; Subasavage et al. 2010); the 0.3 m Perth Exoplanet Survey Telescope in Australia (PEST; 5 transits of 4 targets)[1]; the Danish 1.54 m telescope at La Silla Observatory in Chile (one transit of one target; Andersen et al. 1995); and the Swope 1 m telescope at Las Campanas Observatory in Chile (one transit of one target).

Our methods for carrying out the observations with most of these facilties and reducing the data to light curves have been described in our previous papers (Penev et al. 2013; Mohler-Fischer et al. 2013; Bayliss et al. 2013; Jordán et al. 2014; Hartman et al. 2015; Rabus et al. 2016). The CHAT 0.7 m telescope is a newly commissioned robotic facility at Las Campanas Observatory, built by members of the HATSouth team, and dedicated to the follow-up of transit candidates, especially from HATSouth. The observations from this facility were reduced using the same pipeline that we have applied to the LCO 1 m observations (more description will be provided in Espinoza et al. 2018, in prep). A more detailed description of this facility will be published at a future date (Jordán et al., in preparation).

The time-series photometry data are available in Table 5, and are plotted for each object in Figures 1–10.

### 2.4. *Search for Resolved Stellar Companions*

The Gaia DR2 catalog provides the highest spatial resolution imaging for all of these targets, except HATS-64. Gaia DR2 is sensitive to neighbors with $G \lesssim 20$ mag down to a limiting resolution of $\sim 1''$ (e.g., Ziegler et al. 2018). Table 6 lists the neighbors from Gaia DR2 that are within $10''$ of the planetary systems presented in this paper. For each neighbor we list the separation from the planetary system in arcseconds, and the difference in $G$ magnitude. We also indicate whether the target is potentially a wide binary companion to the planetary host. This latter determination is based on the parallax,

[1] http://pestobservatory.com/

**Table 4.** Summary of spectroscopy observations.

| Instrument | UT Date(s) | # Spec. | Res. $\Delta\lambda/\lambda/1000$ | S/N Range[a] | $\gamma_{RV}$[b] (km s$^{-1}$) | RV Precision[c] (m s$^{-1}$) |
|---|---|---|---|---|---|---|
| **HATS-60** | | | | | | |
| ESO 3.6 m/HARPS | 2017 Apr 23–28 | 5 | 115 | 9–24 | 28.396 | 30 |
| MPG 2.2 m/FEROS | 2017 Jun–Aug | 11[d] | 48 | 32–67 | 28.381 | 22 |
| **HATS-61** | | | | | | |
| MPG 2.2 m/FEROS | 2016 Nov–Dec | 7 | 48 | 32–59 | 54.078 | 26 |
| Euler 1.2 m/Coralie | 2016 Nov 15–18 | 4 | 60 | 12–15 | 54.106 | 67 |
| **HATS-62** | | | | | | |
| ANU 2.3 m/WiFeS | 2012 Apr 10 | 1 | 3 | 88 | ··· | ··· |
| ANU 2.3 m/WiFeS | 2012 Apr 11–13 | 3 | 7 | 20–26 | -12.0 | 4000 |
| AAT 3.9 m/CYCLOPS | 2012 May 8–11 | 3 | 70 | ··· | -10.681 | 110 |
| MPG 2.2 m/FEROS | 2012 May–2013 Sep | 26[d] | 48 | 26–64 | -10.489 | 78 |
| Euler 1.2 m/Coralie | 2012 Jun–Aug | 10[d] | 60 | 11–17 | -10.525 | 69 |
| Magellan 6.5 m/PFS+$I_2$ | 2013 May 20–25 | 8[d] | 76 | ··· | ··· | 32 |
| Magellan 6.5 m/PFS | 2013 May 23 | 2 | 76 | ··· | ··· | ··· |
| VLT UT2 8 m/UVES | 2017 Oct 3–6 | 6 | 60 | 60–63 | -10.5 | ··· |
| **HATS-63** | | | | | | |
| ANU 2.3 m/WiFeS | 2014 Dec 28 | 1 | 3 | 80 | ··· | ··· |
| ANU 2.3 m/WiFeS | 2014 Dec 30–31 | 2 | 7 | 65-102 | -3.1 | 4000 |
| MPG 2.2 m/FEROS | 2017 Jan–Oct | 14[d] | 48 | 27–42 | -4.171 | 44 |
| VLT UT2 8 m/UVES | 2017 Nov 14 | 3 | 60 | 64–67 | -4.2 | ··· |
| **HATS-64** | | | | | | |
| MPG 2.2 m/FEROS | 2013 Nov–2017 Feb | 18[d] | 48 | 45–80 | 7.354 | 70 |
| ANU 2.3 m/WiFeS | 2013 Dec 26 | 1 | 3 | 49 | ··· | ··· |
| ANU 2.3 m/WiFeS | 2013 Dec–2014 Feb | 4 | 7 | 2–73 | 8.0 | 4000 |
| Euler 1.2 m/Coralie | 2014 Mar–2016 Jan | 4[d] | 60 | 18–22 | 7.22 | 490 |
| ESO 3.6 m/HARPS | 2015 Feb–2016 Nov | 13 | 115 | 12–28 | 7.216 | 114 |
| **HATS-65** | | | | | | |
| MPG 2.2 m/FEROS | 2016 Nov–2017 Apr | 6 | 48 | 49–65 | -12.324 | 43 |
| ESO 3.6 m/HARPS | 2016 Nov–2017 Apr | 5[d] | 115 | 17–29 | -12.314 | 29 |
| Euler 1.2 m/Coralie | 2016 Nov 16–17 | 2[d] | 60 | 15–17 | -12.44 | 165 |
| **HATS-66** | | | | | | |
| ANU 2.3 m/WiFeS | 2015 Jan 5 | 1 | 3 | 93 | ··· | ··· |
| ANU 2.3 m/WiFeS | 2015 Oct 3–5 | 2 | 7 | 53–55 | 42.6 | 4000 |
| MPG 2.2 m/FEROS | 2016 Jan–Dec | 13[d] | 48 | 18–44 | 39.940 | 161 |
| VLT UT2 8 m/UVES | 2017 Nov 19 | 6 | 60 | 53–58 | 38.4 | ··· |
| **HATS-67** | | | | | | |
| MPG 2.2 m/FEROS | 2017 Mar–Apr | 13 | 48 | 15–42 | -23.371 | 42 |
| **HATS-68** | | | | | | |
| ESO 3.6 m/HARPS | 2016 Sep–2017 Feb | 4 | 115 | 7–24 | 11.901 | 25 |
| Euler 1.2 m/Coralie | 2016 Sep–Nov | 8 | 60 | 19–31 | 11.836 | 81 |
| MPG 2.2 m/FEROS | 2016 Nov–2017 Oct | 11 | 48 | 26–71 | 11.896 | 40 |
| **HATS-69** | | | | | | |
| ANU 2.3 m/WiFeS | 2014 Oct 4 | 1 | 3 | 63 | ··· | ··· |
| ANU 2.3 m/WiFeS | 2015 Oct 6–7 | 2 | 7 | 38–58 | 0.6 | 4000 |
| MPG 2.2 m/FEROS | 2015 Jul–2017 Jun | 19[d] | 48 | 15–47 | 4.087 | 116 |

[a] S/N per resolution element near 5180 Å. This was not measured for all of the instruments.

[b] For high-precision RV observations included in the orbit determination this is the zero-point RV from the best-fit orbit. For other instruments it is the mean value. We only provide this quantity when applicable.

[c] For high-precision RV observations included in the orbit determination this is the scatter in the RV residuals from the best-fit orbit (which may include astrophysical jitter), for other instruments this is either an estimate of the precision (not including jitter), or the measured standard deviation. We only provide this quantity when applicable.

[d] We list here the total number of spectra collected for each instrument, including observations that were excluded from the analysis due to very low S/N or substantial sky contamination.

**Table 5.** Light curve data for HATS-60–HATS-69.

| Object[a] | BJD[b] (2,400,000+) | Mag[c] | $\sigma_{\rm Mag}$ | Mag(orig)[d] | Filter | Instrument |
|---|---|---|---|---|---|---|
| HATS-60 | 57611.56441 | 12.48078 | 0.00351 | −0.00793 | $r$ | HS/G537.3 |
| HATS-60 | 57721.95234 | 12.49711 | 0.00383 | 0.00840 | $r$ | HS/G537.3 |
| HATS-60 | 57697.02667 | 12.48898 | 0.00387 | 0.00027 | $r$ | HS/G537.3 |
| HATS-60 | 57636.49212 | 12.49794 | 0.00370 | 0.00923 | $r$ | HS/G537.3 |
| HATS-60 | 57615.12701 | 12.48998 | 0.00360 | 0.00127 | $r$ | HS/G537.3 |
| HATS-60 | 57686.34466 | 12.48621 | 0.00366 | −0.00250 | $r$ | HS/G537.3 |
| HATS-60 | 57711.27109 | 12.48277 | 0.00370 | −0.00594 | $r$ | HS/G537.3 |
| HATS-60 | 57586.64113 | 12.47773 | 0.00339 | −0.01098 | $r$ | HS/G537.3 |
| HATS-60 | 57611.56819 | 12.49791 | 0.00343 | 0.00920 | $r$ | HS/G537.3 |
| HATS-60 | 57608.00777 | 12.48966 | 0.00345 | 0.00095 | $r$ | HS/G537.3 |

[a] Either HATS-60, HATS-61, HATS-61, HATS-63, HATS-64, HATS-65, HATS-66, HATS-67, HATS-68 or HATS-69.

[b] Barycentric Julian Date is computed directly from the UTC time without correction for leap seconds.

[c] The out-of-transit level has been subtracted. For observations made with the HATSouth instruments (identified by "HS" in the "Instrument" column) these magnitudes have been corrected for trends using the EPD and TFA procedures applied *prior* to fitting the transit model. This procedure may lead to an artificial dilution in the transit depths. The blend factors for the HATSouth light curves are listed in Table 14. For observations made with follow-up instruments (anything other than "HS" in the "Instrument" column), the magnitudes have been corrected for a quadratic trend in time, and for variations correlated with up to three PSF shape parameters, fit simultaneously with the transit.

[d] Raw magnitude values without correction for the quadratic trend in time, or for trends correlated with the seeing. These are only reported for the follow-up observations.

Note— This table is available in a machine-readable form in the online journal. A portion is shown here for guidance regarding its form and content.

proper motion, and $BP-RP$ color and $G$ magnitude of the neighbor and the planet host. A total of eight neighbors are found within $10''$ of six of the systems, but all of these neighbors are too faint and/or too distant from the planetary host stars to be responsible for the transits or to have any significant impact on the system parameters. HATS-65 has a $5''$ neighbor with a parallax and proper motion that are consistent, within the rather large uncertainties, to those of HATS-65, and with a $BP-RP$ color and $G$ magnitude consistent with falling on the same isochrone. If this is a bound companion it would be an early M dwarf with a mass of $\sim 0.5\,M_{\odot}$ at a projected orbital separation of $2460 \pm 50$ AU from HATS-65. None of the other neighbors identified in Gaia DR2 are compatible with being bound companions to the planetary host stars.

For HATS-64 we also have obtained $z'$-band high-spatial-resolution lucky imaging observations with the Astralux Sur imager (Hippler et al. 2009) on the New Technology Telescope (NTT) on the night of 2015 December 23. The observations were reduced as in Espinoza et al. (2016) and no neighbors were detected. The effective FWHM of the reduced image is $79.10 \pm 5.51$ mas. Figure 11 shows the resulting $5\sigma$ contrast curve. We may exclude neighbors with $\Delta z' < 3$ at $0\farcs2$, and $\Delta z' < 4.8$ at $1''$.

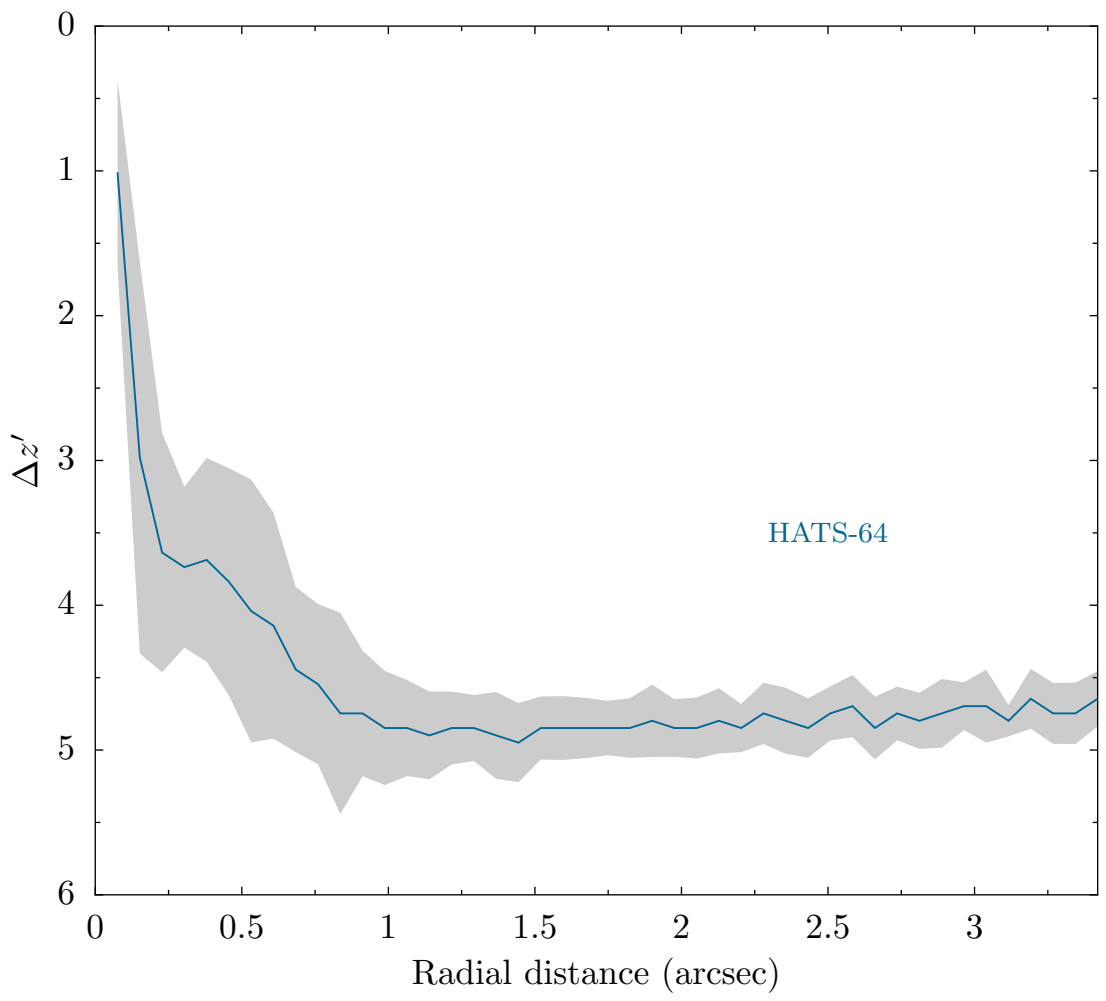

**Figure 11.** $5\sigma$ contrast curve for HATS-64 based on our Astralux Sur $z'$ observation. The gray band shows the variation in the limit in azimuth at a given radius.

**Table 6.** Neighboring Sources in Gaia DR2

| System | Separation $''$ | $\Delta G$ (mag) | Bound Companion? |
|---|---|---|---|
| HATS-61 | 5.63 | 5.98 | no |
| HATS-64 | 7.04 | 6.41 | no |
| HATS-65 | 5.01 | 5.78 | maybe |
| HATS-65 | 8.81 | 3.45 | no |
| HATS-66 | 8.09 | 6.55 | no |
| HATS-67 | 9.76 | 2.59 | no |
| HATS-69 | 7.00 | 6.10 | no |
| HATS-69 | 9.48 | 6.37 | no |

## 3. ANALYSIS

We analyzed the photometric and spectroscopic observations of each system to determine the stellar and planetary parameters, basing our analysis off the methods described in Bakos et al. (2010) and Hartman et al. (2012), but with a number of significant modifications due to the availability of a precise parallax measurement from Gaia DR2. Here we briefly summarize those aspects of the method that have been described in detail elsewhere, and then give a more detailed description of our new modifications.

### 3.1. *Spectroscopic Parameters*

High-precision stellar atmospheric parameters, including $T_{\rm eff\star}$, [Fe/H], $\log g_{\star}$, and $v \sin i$, were measured from the FEROS (HATS-60, HATS-61, HATS-64, HATS-65, HATS-67, HATS-68, and HATS-69) or UVES (HATS-62, HATS-63 and HATS-66) spectra of each target using ZASPE (Brahm et al. 2017b). This code compares the observed high-resolution spectra to a grid of synthetic spectra only in the most sensitive spectral zones, and then uses the systematic differences between the observed spectra and best-fit model to estimate realistic parameter uncertainties.

In our previous work we combined the atmospheric parameters from ZASPE with the stellar density $\rho_{\star}$, determined through modeling the light curves and RV curves, to determine other parameters of the host star, such as its mass, radius, age and luminosity, by comparison with stellar evolution models. In this work we perform the comparison to stellar evolution models simultaneously with the light curve and RV curve fitting, rather than treating these as separate steps. We do, however, continue our practice of performing multiple iterations of the ZASPE analysis. In the first iteration we vary the four above-mentioned parameters. We then perform the joint modeling of the data, described in Section 3.2, which provides an isochrone-based estimate of the stellar surface gravity $\log g_{\star}$. We use this to carry out a second iteration of ZASPE with $\log g_{\star}$ fixed to the value, to determine revised estimates of $T_{\rm eff\star}$, [Fe/H] and $v \sin i$. These revised parameters are then incorporated into a second iteration of the joint modeling to arrive at our final adopted parameters for the system. The spectroscopic parameters measured for HATS-60–HATS-63 are listed, together with catalog astrometry and photometry, in Table 8. Table 9 lists these values for HATS-64–HATS-67, and Table 10 lists the values for HATS-68 and HATS-69.

### 3.2. *Isochrone-based Joint Analysis*

In our previous work we carried out a joint analysis of all available high-precision RVs (fit using a Keplerian orbit) and light curves (fit using a Mandel & Agol 2002 transit model with fixed quadratic limb darkening coefficients from Claret 2004) to measure the stellar density, as well as the orbital and planetary parameters. The fit was performed using a differential evolution Markov Chain Monte Carlo procedure (DEMCMC; ter Braak 2006). In this work we performed a similar analysis for each transiting planet system, but now including the ZASPE $T_{\rm eff\star}$ and [Fe/H] measurements, the Gaia DR2 parallax, and the Gaia DR2 and 2MASS broad-band photometry ($G$, $BP$, $RP$, $J$, $H$ and $K_S$) as observations to be modeled in the fit, together with the RV curve and light curves. The discrepancies between the predicted and measured values for each of these parameters contribute to the overall likelihood computed for a given model. To model these observations we introduce four new model parameters which are allowed to vary in the fit: the distance modulus $(m-M)_0$, the $V$-band extinction $A_V$, and the stellar atmospheric parameters $T_{\rm eff\star}$ and [Fe/H]. Table 7 lists all of the parameters that are varied in the fit, together with the assumed priors. In constructing the likelihood function we assume the observations are independent with Gaussian uncertainties. Each link in the Markov Chain yields a combination of ($T_{\rm eff\star}$, $\rho_\star$, [Fe/H]) which we use to determine the stellar mass, radius, $\log g$, luminosity, and absolute magnitude in the $G$, $BP$, $RP$, $J$, $H$ and $K_S$ bandpasses by comparison with stellar evolution models. Note that $\rho_\star$ is not varied directly in the fit, but rather can be computed from the other transit and orbital parameters which are varied. These absolute magnitudes, together with the model distance modulus and polynomial relations for $A_G(A_V, T_{\rm eff\star})$, $A_{BP}(A_V, T_{\rm eff\star})$, $A_{RP}(A_V, T_{\rm eff\star})$, $A_J(A_V)$, $A_H(A_V)$, and $A_{K_S}(A_V)$ are used to compute model values for the broad-band photometry measurements to be compared to the observations. Here we assume systematic errors of 0.002 mag, 0.005 mag and 0.003 mag on the $G$, $BP$ and $RP$ photometry, respectively, following Evans et al. (2018). These systematic uncertainties are added in quadrature to the statistical uncertainties on the measurements listed in the Gaia DR2 catalog.

We use the PARSEC stellar evolution models (specifically PARSEC release v1.2S + CLIBRI release PR16, as in Marigo et al. 2017) which we generated using the CMD 3.0 web interface by L. Girardi[2]. This differs from our previous work in which we used the Yonsei-Yale ($Y^2$; Yi et al. 2001) models. We chose the PARSEC models because they have incorporated bolometric corrections for the Gaia DR2, SDSS and 2MASS bandpasses. A sequence of isochrones was generated from $\log(t/{\rm yr}) = 6.6$ to $\log(t/{\rm yr}) = 10.13$ in steps of $\Delta(\log(t/{\rm yr})) = 0.05$ for metallicities of $Z$ =0.0001, 0.0002, 0.0005, 0.001, 0.002, 0.004, 0.006, 0.008, 0.009, 0.01, 0.014, 0.015, 0.016, 0.018, 0.02, 0.03, 0.032, 0.036, 0.04, and 0.042, where $Z_\odot = 0.0152$ for these isochrones. We produced a set of isochrones both at $A_V = 0$ and at $A_V = 1$. Given a combination of values ($T_{\rm eff\star}$, $\rho_\star$, [Fe/H]) we generate a model isochrone via trilinear interpolation over these parameters in the tabulated $A_V = 0$ models. We use the same code for this procedure that we have made use of in our previous work with the $Y^2$ isochrones. When a proposed link in the Markov Chain falls outside of the parameter values spanned by the models (e.g., if a star with a density greater than what is allowed by the stellar evolution models at a given temperature and metallicity is proposed) the proposed link is rejected and the previous link is retained. In this manner the fitting procedure used here forces the solutions to match to the theoretical stellar evolution models. We used the $A_V = 1$ models to fit polynomial relations for the extinction in each bandpass as functions of $A_V$ and $T_{\rm eff\star}$.

We assumed uniform priors on the new model parameters $(m-M)_0$, $T_{\rm eff\star}$ and [Fe/H] that we introduced into the fit. For $A_V$ we found that using a uniform prior often led to values that are inconsistent with the expected extinction toward the direction of the source, so we instead made use of the MWDUST 3D Galactic extinction model (Bovy et al. 2016) to tabulate the extinction in 0.1 kpc steps in the direction of the source. For a given $(m-M)_0$ we then perform linear interpolation among these values to estimate the expected $A_V$ at that distance. We treat this expected value as a Gaussian prior, with a $1\sigma$ uncertainty of 0.025 mag for all stars which we found to be the typical discrete change in the predicted $A_V$ when moving toward nearby lines of sight.

### 3.3. *Joint Analysis Using an Empirical Stellar Parameter Method*

[2] http://stev.oapd.inaf.it/cgi-bin/cmd

**Table 7.** Parameters varied in joint analysis

| Parameter | Prior | Notes |
|---|---|---|
| $T_A$ | uniform | mid transit time of first observed transit |
| $T_B$ | uniform | mid transit time of last observed transit |
| $K$ | uniform, $K > 0$ | RV semi-amplitude |
| $\sqrt{e}\cos\omega$ | uniform, $0 \leq e < 1$ | eccentricity parameter, either fixed to zero or varied |
| $\sqrt{e}\sin\omega$ | uniform, $0 \leq e < 1$ | eccentricity parameter, either fixed to zero or varied |
| $R_p/R_\star$ | uniform | ratio of planetary to stellar radius |
| $b^2$ | uniform, $b^2 \geq 0$ | impact parameter squared |
| $\zeta/R_\star$ | uniform | reciprocal of the half duration of the transit |
| $\gamma_i$ | uniform | systemic velocity for RV instrument $i$ |
| $\sigma_{\rm jit,i}$ | $-\log(\sigma_{\rm jit,i})$, $\sigma_{\rm jit,i} > 0$ | jitter for RV instrument $i$ |
| $m_{0,HS,i}$ | uniform | out-of-transit magnitude for HS light curve $i$ |
| $d_{HS,i}$ | uniform, $0 < d_{HS,i} \leq 1$ | transit dilution factor for HS light curve $i$ |
| $m_{0,LC,i}$ | uniform | out-of-transit magnitude for follow-up light curve $i$ |
| $m_{1,LC,i}$ | uniform | linear trend to out-of-transit magnitude for follow-up light curve $i$ |
| $m_{2,LC,i}$ | uniform | quadratic trend to out-of-transit magnitude for follow-up light curve $i$ |
| $S_{0,LC,i}$ | uniform | EPD coefficient for PSF shape parameter $S$ for follow-up light curve $i$ |
| $D_{0,LC,i}$ | uniform | EPD coefficient for PSF shape parameter $S$ for follow-up light curve $i$ |
| $K_{0,LC,i}$ | uniform | EPD coefficient for PSF shape parameter $S$ for follow-up light curve $i$ |
| $d_{\rm mod}$ | $2\ln(\frac{d_{\rm mod}+5}{5}) - \frac{d_{\rm mod}+5}{7650}$ | distance modulus |
| | Gaussian with $\sigma = 0.25$ mag | |
| $A_V$ | mean based on MWDUST model | extinction |
| | $A_V \geq 0$ | |
| $T_{\rm eff}$ | uniform, $T_{\rm eff} > 0$ | host star effective temperature |
| [Fe/H] | uniform | host star metallicity |
| $R_\star$ | $\log R_\star$, $R_\star > 0$ | host star radius, only used for method in Section 3.3, not for method in Section 3.2 |

In addition to the method described above, we also attempted to model the observations of each target using an empirical method for determining the masses and radii of the host stars similar to that proposed by Stassun et al. (2018). This method makes use of the Gaia DR2 parallax, the broad-band photometry and the spectroscopically determined $T_{\rm eff\star}$ to directly determine the radius of the star, and then combines this with the density constrained from the transits to directly determine the mass of the star. We applied this method by following a similar procedure to that detailed above, except that instead of comparing a given proposed combination of ($T_{\rm eff\star}$, $\rho_\star$, [Fe/H]) to the theoretical stellar evolution models, we instead introduce the stellar radius as a new free parameter in the model (adopting a uniform prior on $\log R_\star$) and used a combination of ($T_{\rm eff\star}$, $\log g_\star$, [Fe/H]) to determine the bolometric correction (reverse engineered from the PARSEC models) to apply to the bolometric magnitude to model the observed magnitude in each bandpass. This method has the benefit that it does not force the parameters of the system to agree with the theoretical stellar evolution models, which may have undetermined systematic errors, but in practice we found that for many of the systems it leads to a poor constraint on the stellar mass spanning a wide parameter range that is certainly unphysical. This is demonstrated in Table 17 where we compare the stellar masses and radii inferred for each system using the isochrone and empirical models. While the radii from both methods are comparable, the masses from the empirical modeling have uncertainties that are typically an order of magnitude larger than the mass uncertainties from the isochrone-based method.

### 3.4. *Adopted Parameters and Comparisons to Observations*

Figures 1–10 include comparisons between the broadband photometric measurements of each system and the models from our isochrone-based analysis. The plots include absolute $G$ magnitude vs. de-reddened $BP - RP$ color, and observed broad-band spectral energy distributions (SEDs). We find that the Gaia photometry and parallaxes, and the 2MASS photometry are consistent with the models for all ten systems.

Our final set of adopted stellar parameters derived from this analysis are listed in Table 11 for HATS-60–HATS-63, in Table 12 for HATS-64–HATS-67, and in Table 13 for HATS-68 and HATS-69. The parameters listed here are from the isochrone-based analysis (Section 3.2) which we adopt for the remainder of the paper. Ordered from least to most massive, the stars have masses and radii of:

| | | |
|---|---|---|
| HATS-69 | $0.892^{+0.011}_{-0.016}\,M_{\odot}$ | $0.8785 \pm 0.0077\,R_{\odot}$ |
| HATS-62 | $0.896^{+0.015}_{-0.010}\,M_{\odot}$ | $0.933^{+0.019}_{-0.013}\,R_{\odot}$ |
| HATS-63 | $0.931 \pm 0.019\,M_{\odot}$ | $1.070 \pm 0.012\,R_{\odot}$ |
| HATS-61 | $1.076 \pm 0.014\,M_{\odot}$ | $1.664 \pm 0.024\,R_{\odot}$ |
| HATS-60 | $1.097^{+0.010}_{-0.016}\,M_{\odot}$ | $1.460 \pm 0.024\,R_{\odot}$ |
| HATS-65 | $1.257 \pm 0.028\,M_{\odot}$ | $1.310 \pm 0.027\,R_{\odot}$ |
| HATS-68 | $1.351 \pm 0.014\,M_{\odot}$ | $1.748 \pm 0.026\,R_{\odot}$ |
| HATS-66 | $1.411 \pm 0.022\,M_{\odot}$ | $1.841 \pm 0.041\,R_{\odot}$ |
| HATS-67 | $1.435 \pm 0.021\,M_{\odot}$ | $1.441 \pm 0.026\,R_{\odot}$ |
| HATS-64 | $1.564 \pm 0.028\,M_{\odot}$ | $2.113 \pm 0.071\,R_{\odot}$ |

Our final set of adopted planetary parameters derived from the isochrone-based method are listed in Table 14 for HATS-60b–HATS-63b, in Table 15 for HATS-64b–HATS-67b, and in Table 16 for HATS-68b and HATS-69b. We considered both models where the eccentricity of the planetary orbit was allowed to vary, and models where it was fixed to zero. We find that for all ten systems the observations are consistent with zero eccentricity, and we adopt the fixed circular orbit solutions. We list the 95% confidence upper limit on the eccentricity for each planet.

For two of the transiting planets (HATS-62b and HATS-69b) the measured orbital semi-amplitudes are not detected with at least $3\sigma$ confidence. For HATS-62b we measure $K = 10.2 \pm 7.8\,\mathrm{m\,s^{-1}}$, leading to $M_p = 0.070 \pm 0.053\,M_{\mathrm{J}}$, while for HATS-69b we measure $K = 52 \pm 28\,\mathrm{m\,s^{-1}}$, leading to $M_p = 0.31 \pm 0.17\,M_{\mathrm{J}}$. For these two planets we list the 95% confidence upper limits on their masses of $M_p < 0.179\,M_{\mathrm{J}}$ and $M_p < 0.577\,M_{\mathrm{J}}$, for HATS-62b and HATS-69b, respectively (if we exclude the outlier FEROS observation of HATS-62 seen in Figure 3 from the fit, both the best-estimate and upper limit on the planet mass would be lower by 10%). With respective radii of $1.055 \pm 0.025\,R_{\mathrm{J}}$ and $0.945 \pm 0.022\,R_{\mathrm{J}}$, these two planets also have the smallest radii among the sample of planets presented in this paper. Based on their equilibrium temperatures, radii, and mass limits, we conclude that HATS-62b is likely an inflated hot Super-Neptune while HATS-69b may be a hot Saturn.

Ordered from least to most massive, the eight other planets have masses and radii of:

| | | |
|---|---|---|
| HATS-60b | $0.662 \pm 0.055\,M_{\mathrm{J}}$ | $1.153 \pm 0.053\,R_{\mathrm{J}}$ |
| HATS-65b | $0.821 \pm 0.083\,M_{\mathrm{J}}$ | $1.501 \pm 0.050\,R_{\mathrm{J}}$ |
| HATS-64b | $0.96 \pm 0.20\,M_{\mathrm{J}}$ | $1.679 \pm 0.081\,R_{\mathrm{J}}$ |
| HATS-63b | $0.96 \pm 0.12\,M_{\mathrm{J}}$ | $1.207 \pm 0.039\,R_{\mathrm{J}}$ |
| HATS-68b | $1.290 \pm 0.059\,M_{\mathrm{J}}$ | $1.232^{+0.039}_{-0.029}\,R_{\mathrm{J}}$ |
| HATS-67b | $1.45 \pm 0.12\,M_{\mathrm{J}}$ | $1.685 \pm 0.047\,R_{\mathrm{J}}$ |
| HATS-61b | $3.40 \pm 0.14\,M_{\mathrm{J}}$ | $1.195 \pm 0.067\,R_{\mathrm{J}}$ |
| HATS-66b | $5.33 \pm 0.68\,M_{\mathrm{J}}$ | $1.411 \pm 0.084\,R_{\mathrm{J}}$ |

One interesting result of combining the Gaia DR2 observations and the PARSEC stellar evolution models directly into the joint analysis of the data is that the stellar density and orbital inclination are much more tightly constrained than they are from the light curves alone. For example, for HATS-61b we find an inclination of $87.15 \pm 0.18$ and stellar density of $0.330 \pm 0.014\,\mathrm{g\,cm^{-3}}$, compared to values of $86.93^{+0.59}_{-0.94}$ and $0.308^{+0.052}_{-0.074}\,\mathrm{g\,cm^{-3}}$ based on the empirical model. The uncertainties for the reciprocal half transit duration, by contrast, are nearly identical between the two methods with $\zeta/R_{\star} = 9.54 \pm 0.13\,\mathrm{d^{-1}}$ for the isochrone-based method and $\zeta/R_{\star} = 9.53 \pm 0.13\,\mathrm{d^{-1}}$ for the empirical method. What is happening is that the tight constraint on the stellar radius, stemming from the Gaia DR2 measurements, when combined with the effective temperature and metallicity, and coupled with the stellar evolution models, forces a tight constraint on the stellar mass, which in turn leads to a tighter constraint on the bulk stellar density than is measured from the light curves. This, together with the well-measured value of $\zeta/R_{\star}$, leads to a tight constraint on the inclination. It is important to note here that the uncertainties that we have derived for these systems do not include possible systematic errors in the stellar models. If these errors exceed the listed uncertainties, then the errors on most of the inferred planet and stellar parameters would be larger as well.

### 3.5. *Blend Analysis*

In order to rule out the possibility that any of these objects is a blended stellar eclipsing binary system, we carried out a blend analysis of the photometric data following Hartman et al. (2012). As for the joint analysis of the data described in Sections 3.2 and 3.3, we had to modify the procedure to account for the Gaia DR2 measurements. These modifications include incorporating the parallax and Gaia DR2 $G$, $BP$ and $RP$ broad-band photometry into the fit, using the PARSEC stellar evolution models (Marigo et al. 2017) in place of the older Padova models from Girardi et al. (2000), and using the MWDUST 3D Galactic extinction model (Bovy et al. 2016) to place a prior on $A_V$ as we did in the joint analysis. We find that largely thanks to the strong constraint on the distance to the brightest source from the Gaia DR2 parallax, we can easily rule out blended stellar eclipsing binary models for all ten objects.

Table 18 lists, for each system, the $\chi^2$ difference between the best-fit blend models and the best-fit single star with a planet model (referred to as the H-p model) for three different blend model scenarios. The scenarios, which we label H,S-s, H,S-s$_{\rm BGEB}$, and H-p,s following the nomenclature from Hartman et al. (2009), correspond to a hierarchical triple star system where the two fainter stars form an eclipsing binary, a blend between a bright foreground star and a fainter background eclipsing binary star system, and a bright star with a transiting planet and a fainter unresolved stellar companion. For each case we list both the total $\Delta\chi^2$, and the contribution to $\Delta\chi^2$ from the Gaia DR2 parallax $\varpi$. We also list the mass $M_3$ of the unresolved binary companion for the best-fit H-p,s model, together with the 95% confidence ($\Delta\chi^2 = 3.84$) upper limits on the mass and luminosity ratio $L_3/L_1$ for any binary companion.

We find that both the H,S-s and H,S-s$_{\rm BGEB}$ blend scenarios can be rejected for most of the systems based on their fit to the light curves, broad-band photometry, parallax and atmospheric parameters. In most cases the Gaia DR2 parallax provides a significant contribution to the total $\Delta\chi^2$. In these cases the combined light from any blend of stars capable of fitting the photometry requires the brightest source to be at a greater distance than is measured. In three cases (HATS-60, -63 and -68) the blended eclipsing binary models cannot be rejected with at least $5\sigma$ confidence based on the abovementioned observations. In these cases we are able to reject the blended eclipsing binary scenarios based on their inability to reproduce the observed RV and/or BS variations. We arrive at this conclusion by simulating the expected spectroscopic CCF of each blend model that we tested, and using this to estimate the expected variation in the RVs and BS values. We note that blended eclipsing binary models are also inconsistent with the observed RV variation and/or lack of BS variations for the other seven systems as well. We conclude that all ten systems contain transiting planets, and that none of them are blended stellar eclipsing binary objects.

While we are able to rule out blended stellar eclipsing binary scenarios for all ten of the systems, we are not able to rule out the H-p,s scenario (i.e., a transiting planet system with an additional unresolved stellar companion) for any of these systems. In fact, for three of the objects (HATS-62, -64 and -65) the H-p,s scenario provides a sufficient improvement to $\chi^2$ to suggest that unresolved stellar companions may be present. For HATS-62 the best-fit model has a companion of mass $0.45\,M_\odot$ leading to an improvement in $\chi^2$ of 16.3. In this case the planet host has a mass and radius that are larger by 0.12% and 0.70%, respectively, while the planet has a radius that is larger by 1.7%. The RVs would be slightly diluted as well, leading to an underestimate of the planetary mass at a similar fractional level. For HATS-64 the best-fit model has a companion of mass $1.01\,M_\odot$ which improves $\chi^2$ by 8.1. In this case the host has a mass that is larger by 0.32% and the radius is smaller by 2.3%. The planet has a radius that is larger by 9.0%. Note that in this case the companion star has $\Delta z = 2.3$ with respect to the host star. Given the limits on resolved companions based on our Astralux observations of this star (Figure 11), such a companion would need to be within $\sim 0\farcs1$ of the host star, or $\sim 100$ AU. For HATS-65 the best-fit model has a companion of mass $0.53\,M_\odot$ leading to a reduction in $\chi^2$ of 11.4. The planet host has a mass and radius that are smaller by 0.12% and 1.3%, respectively. The planet would have a radius that is smaller by 2.4%. Although the best-fit H-p,s model yields a transit depth that is $\sim 0.6$ mmag shallower than the H-p model, this appears to be balanced by changes to the systematic trend model so that the resulting model provides an indistinguishable fit to the light curves. Given the relatively modest improvement in $\chi^2$, and the possibility of systematic uncertainties in the predicted Gaia DR2 broad-band photometry from our interpolation of the PARSEC isochrones, we do not claim definitive evidence for stellar companions in these systems. Clear evidence for companions could be obtained through higher spatial resolution imaging, continued RV follow-up, or perhaps through astrometric variations detected in future Gaia data releases.

For all systems we place upper limits on the mass and luminosity ratio of any unresolved stellar companion. Except for HATS-61 and HATS-64, we are able to exclude companions with luminosities greater than 10% that of the planetary host, and in some cases we can even exclude companions with luminosities as low as 1% that of the host. The estimated physical parameters of the planet and host would thus be only modestly affected if a stellar companion were detected. For HATS-61 and HATS-64 brighter companions could still be present, po-tentially leading to substantial changes in the estimated parameters.

## 4. DISCUSSION

We have presented the discovery of ten new transiting planet systems from HATSouth. The planets are shown on mass–radius, equilibrium temperature–radius, and semimajor axis–mass diagrams in Figure 12. We compare the newly discovered planets to the sample of previously discovered transiting planets as listed in the NASA Exoplanet Archive as of 2018 Aug 8. The newly discovered planets follow the well-established trends, though some are expanding slightly the envelopes of points in these diagrams.

With a mass of $0.070 \pm 0.053\,M_{\rm J}$ ($< 0.179\,M_{\rm J}$ 95% confidence upper limit) and a radius of $1.055 \pm 0.025\,R_{\rm J}$, HATS-62b is the largest radius Super-Neptune found to date. The next two least massive planets known with radii larger than Jupiter are WASP-127 ($M_p = 0.18 \pm 0.02\,M_{\rm J}$, $R_p = 1.37 \pm 0.04\,R_{\rm J}$; Lam et al. 2017) and KELT-11 ($M_p = 0.195 \pm 0.019\,M_{\rm J}$, $R_p = 1.37 \pm 0.15\,R_{\rm J}$; Pepper et al. 2017). It is perhaps not a coincidence that this large radius Super-Neptune is also located near the lower envelope of close-in gas giant planets in the semimajor axis–mass diagram shown in Figure 12. This envelope marks the upper edge of the sub-Jovian desert (e.g., Mazeh et al. 2016), and may trace the tidal disruption limit of gas giants undergoing high-eccentricity migration (Owen & Lai 2018). The probable hot Saturn HATS-69b also lies along this boundary in the semimajor axis–mass diagram, though its radius is not exceptional for its mass.

With an equilibrium temperature of $2193 \pm 22$ K, HATS-67b is among the most highly irradiated hot Jupiters known. Not surprisingly, it is also highly inflated with a radius of $1.685 \pm 0.047\,R_{\rm J}$.

The planet HATS-66b is a massive Super-Jupiter with $M_p = 5.33 \pm 0.68\,M_{\rm J}$. Such planets are relatively rare. There are only 20 transiting planets listed in the NASA Exoplanet Archive more massive than HATS-66b. Recently Schlaufman (2018) has argued, based on the absence of a correlation between occurrence and host star metallicity, that objects with $M_p > 4\,M_{\rm J}$ may have formed through disk instability rather than core accretion. Objects with $4\,M_{\rm J} < M_p < 10\,M_{\rm J}$ may thus be more related to brown dwarfs than to planets. HATS-66b orbits a solar-metalicity star with [Fe/H]$= 0.000 \pm 0.044$. HATS-66b has a rather large radius of $1.411 \pm 0.084\,R_{\rm J}$ for a planet of its mass, which is in line with its high equilibrium temperature of $1998 \pm 21$ K.

The planet HATS-61b is a $M_p = 3.40 \pm 0.14\,M_{\rm J}$ Super-Jupiter on a relatively long period orbit of $7.817953 \pm 0.000024$ days. This is the second longest period planet announced so far by HATSouth. The host star is relatively old ($8.90^{+0.31}_{-0.41}$ Gyr) and beginning to evolve off the main sequence, with a current luminosity that is $\sim 2.7$ times greater than what it would have been at the zero-age main sequence (ZAMS). Thus, despite its relatively long period, the planet is expected to be hot with an estimated equilibrium temperature of $1226.1 \pm 7.3$ K. The $1.195 \pm 0.067\,R_{\rm J}$ radius of the planet is consistent with the observed equilibrium-temperature–radius correlation (e.g., the empirical relation from Enoch et al. 2012, yields a predicted radius of $1.12 \pm 0.11\,R_{\rm J}$). If the equilibrium temperature were adjusted to the expected value at ZAMS (assuming the same semimajor axis), the radius of the planet would be near the upper boundary in the equilibrium-temperature–radius relation (the predicted radius based on the Enoch et al. 2012 relation would be $1.02 \pm 0.11\,R_{\rm J}$). HATS-61b is potentially a re-inflated super-Jupiter which is dynamically increasing in size as its host becomes more luminous (Lopez & Fortney 2016; Grunblatt et al. 2016; Hartman et al. 2016). However, given the intrinsic scatter in planetary radius at fixed temperature and mass, this conclusion is by no means definitive.

The planet discoveries presented here are among the first discoveries from HATSouth to take advantage of the high-precision parallax measurements provided by Gaia DR2. This has enabled much more precise characterizations of the planetary host stars than would be possible otherwise. For the 10 systems presented in this work, the median relative precision of the stellar radius is 1.9% (c.f., 5.6% for previous HATSouth discoveries that did not incorporate Gaia DR2 into their analyses), the median relative precision of the stellar mass is 1.7% (c.f., 3.9% for previous HATSouth discoveries), and the median relative precision of the planetary radius is 3.5% (c.f., 6.6% for previous HATSouth discoveries). The precision of the planetary masses, however, is still limited by the RV observations.

In order to make use of the Gaia DR2 observations we have made a number of significant modifications to our analysis procedures. These include incorporating the stellar isochrone look-up directly into the Markov Chain Monte Carlo joint modeling of the transiting planet observations, applying a prior on the interstellar extinction using a 3D Galactic dust model, and making use of the PARSEC stellar models in place of the older YY models. We have also tried to apply the purely empirical stellar modelling procedure of Stassun et al. (2018) to the data, but find that our constraints on the stellar density are too poor, given the present ground-based photometry, to provide a reasonably precise determination of the stellar masses.

The Gaia DR2 observations also allow us to identify three systems (HATS-62, -64 and -65) as showing suggestive evidence for the presence of an unresolved binary star companion to the planetary host star. Additional high-resolution imaging, and long-term RV monitoring would be needed to confirm these companions if they are present.

Nine of the ten planets presented here are expected to be observed by the NASA *TESS* mission during the upcoming year. It is unknown at this time which, if any, of these systems will be observed at two minute cadence, but any object within the field of view will at least be observed through the full frame images. These data will enable more precise measurements of their orbital inclinations, stellar densities, and planetary radii, and may enable the discovery of additional transiting planets in these systems. It may also be possible to measure photometric rotation periods for the host stars if they are active, and if they have periods that are shorter than the timespan of the observations. The only system that will not be observed by *TESS* is HATS-65, which is at ecliptic coordinates $\lambda = 291.07°$, $\beta = -7.96°$, and will likely fall in the gap between sectors 1 and 13 of the primary mission. Two of the planets (HATS-62 and HATS-68) are located within sector 1 of the mission, which is currently being observed at the time of writing. HATS-61, HATS-66 and HATS-68 are expected to be observed in two sectors, and receive 54 days of continuous coverage, while the other systems will be observed in only one sector, and receive 27 days of coverage.

We thank the anonymous referee for their careful review of our paper, which has significantly improved its quality. Development of the HATSouth project was funded by NSF MRI grant NSF/AST-0723074, operations have been supported by NASA grants NNX09AB29G, NNX12AH91H, and NNX17AB61G, and follow-up observations have received partial support from grant NSF/AST-1108686. A.J. acknowledges support from FONDECYT project 1171208, BASAL CATA PFB-06, and project IC120009 "Millennium Institute of Astrophysics (MAS)" of the Millenium Science Initiative, Chilean Ministry of Economy. N.E. is supported by CONICYT-PCHA/Doctorado Nacional. R.B. acknowledges support from FONDECYT Post-doctoral Fellowship Project No. 3180246. N.E. acknowledges support from project IC120009 "Millenium Institute of Astrophysics (MAS)" of the Millennium Science Initiative, Chilean Ministry of Economy. L.M. acknowledges support from the Italian Minister of Instruction, University and Research (MIUR) through FFABR 2017 fund. L.M. acknowledges support from the University of Rome Tor Vergata through "Mission: Sustainability 2016" fund. V.S. acknowledges support form BASAL CATA PFB-06. A.V. is supported by the NSF Graduate Research Fellowship, Grant No. DGE 1144152. This work is based on observations made with ESO Telescopes at the La Silla Observatory. This paper also makes use of observations from the LCOGT network. Some of this time was awarded by NOAO. We acknowledge the use of the AAVSO Photometric All-Sky Survey (APASS), funded by the Robert Martin Ayers Sciences Fund, and the SIMBAD database, operated at CDS, Strasbourg, France. Operations at the MPG 2.2 m Telescope are jointly performed by the Max Planck Gesellschaft and the European Southern Observatory. We thank the MPG 2.2m telescope support team for their technical assistance during observations. This work has made use of data from the European Space Agency (ESA) mission *Gaia* (https://www.cosmos.esa.int/gaia), processed by the *Gaia* Data Processing and Analysis Consortium (DPAC, https://www.cosmos.esa.int/web/gaia/dpac/consorti Funding for the DPAC has been provided by national institutions, in particular the institutions participating in the *Gaia* Multilateral Agreement. This research has made use of the NASA Exoplanet Archive, which is operated by the California Institute of Technology, under contract with the National Aeronautics and Space Administration under the Exoplanet Exploration Program.

*Facilities:* HATSouth, LCOGT, FTS, CTIO:0.9m, Danish 1.54m Telescope (DFOSC), Swope, Max Planck:2.2m (FEROS), ESO:3.6m (HARPS), Euler1.2m (Coralie), ATT (WiFeS), AAT (CYCLOPS), Magellan:Clay (PFS), VLT:Kueyen (UVES), NTT (Astralux Sur), Gaia

*Software:* ZASPE (Brahm et al. 2017b), CERES (Brahm et al. 2017a), FITSH (Pál 2012), VARTOOLS (Hartman, & Bakos 2016), BLENDANAL (Hartman et al. 2011), PARSEC (Marigo et al. 2017), LCOGTDD (Espinoza 2018, https://github.com/nespinoza/lcogtDD), astropy (Astropy Collaboration et al. 2018)

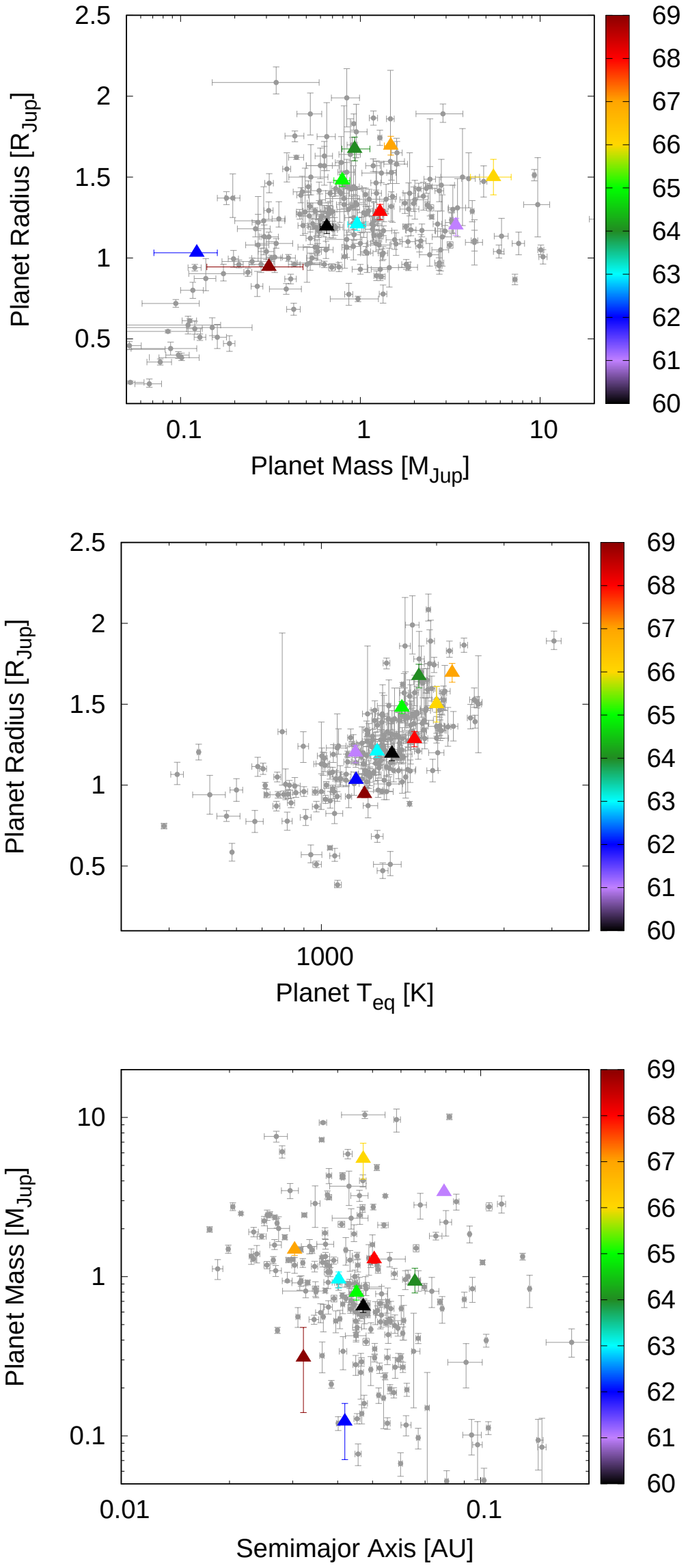

**Figure 12.** The ten newly discovered transiting planets are shown on mass–radius (top), equilibrium temperature–radius (middle), and semimajor axis–mass (bottom) diagrams. In each case the colored points represent the newly discovered planets with the color of the point designating the HATS planet number, as indicated in the color-bars on the right-hand-side of each plot. The grayscale points show other transiting planets listed in the NASA Exoplanet Archive as of 2018 Aug 8. We only show planets with definite mass measurements and with nonzero values for the semimajor axis and equilibrium temperature in the database. We also exclude planets with large uncertainties on their equilibrium temperature or semimajor axis. In the middle panel we only show planets with measured masses greater than 0.1 $M_{\rm J}$. HATS-62b stands out in the mass–radius diagram as an object that is located along the upper envelope of points. It is the least massive planet discovered to date with a radius larger than that of Jupiter. HATS-62b and HATS-69b stand out in the semimajor axis–mass diagram as being located along the lower envelope of points delineating the so-called sub-Jovian desert (e.g., Mazeh et al. 2016).

**Table 8.** Astrometric, Spectroscopic and Photometric parameters for HATS-60, HATS-61, HATS-62 and HATS-63

| Parameter | **HATS-60** Value | **HATS-61** Value | **HATS-62** Value | **HATS-63** Value | Source |
|---|---|---|---|---|---|
| Astrometric properties and cross-identifications | | | | | |
| 2MASS-ID | 22452736-1459303 | 04063786-2520589 | 20494783-2418124 | 04294044-2811501 | |
| TIC-ID | 145750719 | 44745133 | 336732544 | 178879588 | |
| GAIA DR2-ID | 2596986648798061952 | 4890849134501995392 | 6806639397331208320 | 4891362198412001408 | |
| R.A. (J2000) | $22^{h}45^{m}27.3643s$ | $04^{h}06^{m}37.8676s$ | $20^{h}49^{m}47.8333s$ | $04^{h}29^{m}40.4529s$ | GAIA DR2 |
| Dec. (J2000) | $-14^{\circ}59'30.3457''$ | $-25^{\circ}20'58.9560''$ | $-24^{\circ}18'12.4965''$ | $-28^{\circ}11'50.2340''$ | GAIA DR2 |
| $\mu_{\rm R.A.}$ (mas yr$^{-1}$) | $3.481 \pm 0.067$ | $3.997 \pm 0.022$ | $0.489 \pm 0.054$ | $5.777 \pm 0.022$ | GAIA DR2 |
| $\mu_{\rm Dec.}$ (mas yr$^{-1}$) | $-2.787 \pm 0.052$ | $9.892 \pm 0.032$ | $-8.074 \pm 0.035$ | $16.810 \pm 0.029$ | GAIA DR2 |
| parallax (mas) | $2.027 \pm 0.035$ | $1.442 \pm 0.018$ | $1.884 \pm 0.037$ | $1.576 \pm 0.015$ | GAIA DR2 |
| Spectroscopic properties | | | | | |
| $T_{\rm eff\star}$ (K) | $5698 \pm 58$ | $5630 \pm 71$ | $5536 \pm 33$ | $5637 \pm 46$ | ZASPE[a] |
| [Fe/H] | $0.320 \pm 0.028$ | $0.220 \pm 0.043$ | $0.120 \pm 0.024$ | $0.060 \pm 0.040$ | ZASPE |
| $v \sin i$ (km s$^{-1}$) | $3.84 \pm 0.43$ | $3.52 \pm 0.42$ | $0.50 \pm 0.27$ | $1.77 \pm 0.45$ | ZASPE |
| $v_{\rm mac}$ (km s$^{-1}$) | $3.869 \pm 0.088$ | $3.76 \pm 0.11$ | $3.620 \pm 0.050$ | $3.775 \pm 0.070$ | Assumed[b] |
| $v_{\rm mic}$ (km s$^{-1}$) | $1.036 \pm 0.032$ | $0.999 \pm 0.037$ | $0.952 \pm 0.016$ | $1.003 \pm 0.024$ | Assumed[b] |
| $\gamma_{\rm RV}$ (m s$^{-1}$) | $28379.9 \pm 6.7$ | $54079 \pm 14$ | $-10489 \pm 13$ | $-4171 \pm 13$ | FEROS[c] |
| Photometric properties | | | | | |
| $G$ (mag)[d] | $12.50040 \pm 0.00030$ | $13.06820 \pm 0.00030$ | $13.83210 \pm 0.00030$ | $13.72130 \pm 0.00020$ | GAIA DR2 |
| $BP$ (mag)[d] | $12.8894 \pm 0.0024$ | $13.4659 \pm 0.0012$ | $14.2608 \pm 0.0012$ | $14.0919 \pm 0.0014$ | GAIA DR2 |
| $RP$ (mag)[d] | $11.9719 \pm 0.0017$ | $12.52090 \pm 0.00060$ | $13.2562 \pm 0.0012$ | $13.2011 \pm 0.0011$ | GAIA DR2 |
| $B$ (mag) | $13.394 \pm 0.023$ | $14.020 \pm 0.036$ | $14.862 \pm 0.027$ | $14.595 \pm 0.049$ | APASS[e] |
| $V$ (mag) | $12.641 \pm 0.028$ | $13.233 \pm 0.024$ | $14.015 \pm 0.037$ | $13.8920 \pm 0.0090$ | APASS[e] |
| $g$ (mag) | $12.987 \pm 0.030$ | $13.593 \pm 0.046$ | $14.421 \pm 0.058$ | $14.183 \pm 0.057$ | APASS[e] |
| $r$ (mag) | $12.439 \pm 0.040$ | $13.014 \pm 0.021$ | $13.776 \pm 0.019$ | $13.674 \pm 0.024$ | APASS[e] |
| $i$ (mag) | $12.288 \pm 0.046$ | $12.854 \pm 0.056$ | $13.591 \pm 0.011$ | $13.482 \pm 0.032$ | APASS[e] |
| $J$ (mag) | $11.377 \pm 0.023$ | $11.875 \pm 0.028$ | $12.573 \pm 0.021$ | $12.631 \pm 0.024$ | 2MASS |
| $H$ (mag) | $11.070 \pm 0.022$ | $11.565 \pm 0.024$ | $12.196 \pm 0.027$ | $12.290 \pm 0.025$ | 2MASS |
| $K_s$ (mag) | $10.988 \pm 0.023$ | $11.478 \pm 0.025$ | $12.109 \pm 0.026$ | $12.216 \pm 0.024$ | 2MASS |

[a] ZASPE = Zonal Atmospherical Stellar Parameter Estimator routine for the analysis of high-resolution spectra (Brahm et al. 2017b), applied to the FEROS or UVES spectra of each system. These parameters rely primarily on ZASPE, but have a small dependence also on the iterative analysis incorporating the isochrone search and global modeling of the data.

[b] The macro and microturbulence parameters adopted in a given iteration of ZASPE are calculated from the trial effective temperature using the polynomial relations given in Brahm et al. (2017b). The uncertainties listed here on these parameters give the scatter in the adopted values propagated from the uncertainty on the effective temperature and do not include the uncertainty in the assumed polynomial relations themselves.

[c] The error on $\gamma_{\rm RV}$ is determined from the orbital fit to the RV measurements, and does not include the systematic uncertainty in transforming the velocities to the IAU standard system. The velocities have not been corrected for gravitational redshifts.

[d] The listed uncertainties for the Gaia DR2 photometry are taken from the catalog. For the analysis we assume additional systematic uncertainties of 0.002 mag, 0.005 mag and 0.003 mag for the G, BP and RP bands, respectively.

[e] From APASS DR6 for as listed in the UCAC 4 catalog (Zacharias et al. 2013).

**Table 9.** Astrometric, Spectroscopic and Photometric parameters for HATS-64, HATS-65, HATS-66 and HATS-67

| | **HATS-64** | **HATS-65** | **HATS-66** | **HATS-67** | |
|---|---|---|---|---|---|
| Parameter | Value | Value | Value | Value | Source |
| Astrometric properties and cross-identifications | | | | | |
| 2MASS-ID | 09370902-2948015 | 19314555-2644246 | 06453475-3352540 | 12005011-4608110 | |
| TIC-ID | 189625051 | 169504920 | 52689469 | 272212970 | |
| GAIA DR2-ID | 5632704511826797824 | 6766134630213144704 | 5582647836223843840 | 6144060260072337024 | |
| R.A. (J2000) | $09^{\mathrm{h}}37^{\mathrm{m}}09.0299\mathrm{s}$ | $19^{\mathrm{h}}31^{\mathrm{m}}45.5518\mathrm{s}$ | $06^{\mathrm{h}}45^{\mathrm{m}}34.7574\mathrm{s}$ | $12^{\mathrm{h}}00^{\mathrm{m}}50.1183\mathrm{s}$ | GAIA DR2 |
| Dec. (J2000) | $-29°48'01.5746''$ | $-26°44'24.7250''$ | $-33°52'54.1300''$ | $-46°08'11.1247''$ | GAIA DR2 |
| $\mu_{\mathrm{R.A.}}$ (mas yr$^{-1}$) | $-3.127 \pm 0.070$ | $-3.495 \pm 0.081$ | $-3.369 \pm 0.026$ | $-6.082 \pm 0.030$ | GAIA DR2 |
| $\mu_{\mathrm{Dec.}}$ (mas yr$^{-1}$) | $-1.527 \pm 0.067$ | $-0.158 \pm 0.076$ | $2.549 \pm 0.029$ | $0.801 \pm 0.022$ | GAIA DR2 |
| parallax (mas) | $0.897 \pm 0.035$ | $2.000 \pm 0.050$ | $0.648 \pm 0.016$ | $1.013 \pm 0.025$ | GAIA DR2 |
| Spectroscopic properties | | | | | |
| $T_{\mathrm{eff}\star}$ (K) | $6635 \pm 85$ | $6660 \pm 110$ | $6500 \pm 78$ | $6570 \pm 100$ | ZASPE |
| [Fe/H] | $0.220 \pm 0.042$ | $0.180 \pm 0.062$ | $0.000 \pm 0.044$ | $0.380 \pm 0.056$ | ZASPE |
| $v \sin i$ (km s$^{-1}$) | $12.65 \pm 0.22$ | $7.83 \pm 0.30$ | $12.86 \pm 0.17$ | $5.19 \pm 0.38$ | ZASPE |
| $v_{\mathrm{mac}}$ (km s$^{-1}$) | $5.31 \pm 0.13$ | $5.34 \pm 0.16$ | $5.10 \pm 0.19$ | $5.20 \pm 0.15$ | Assumed |
| $v_{\mathrm{mic}}$ (km s$^{-1}$) | $1.96 \pm 0.14$ | $2.00 \pm 0.18$ | $1.76 \pm 0.11$ | $1.85 \pm 0.15$ | Assumed |
| $\gamma_{\mathrm{RV}}$ (m s$^{-1}$) | $7358 \pm 16$ | $-12318 \pm 12$ | $39938 \pm 58$ | $-23368 \pm 13$ | FEROS |
| Photometric properties | | | | | |
| $G$ (mag) | $12.83900 \pm 0.00020$ | $12.37930 \pm 0.00040$ | $14.00860 \pm 0.00030$ | $13.55960 \pm 0.00020$ | GAIA DR2 |
| $BP$ (mag) | $13.0978 \pm 0.0014$ | $12.6817 \pm 0.0020$ | $14.2831 \pm 0.0012$ | $13.8512 \pm 0.0013$ | GAIA DR2 |
| $RP$ (mag) | $12.4265 \pm 0.0015$ | $11.92130 \pm 0.00080$ | $13.57490 \pm 0.00090$ | $13.11670 \pm 0.00070$ | GAIA DR2 |
| $B$ (mag) | $13.416 \pm 0.030$ | $13.067 \pm 0.020$ | $14.630 \pm 0.030$ | $14.207 \pm 0.030$ | APASS |
| $V$ (mag) | $12.924 \pm 0.030$ | $12.497 \pm 0.020$ | $14.095 \pm 0.030$ | $13.653 \pm 0.010$ | APASS |
| $g$ (mag) | $13.130 \pm 0.030$ | $12.747 \pm 0.030$ | $14.344 \pm 0.010$ | $13.882 \pm 0.020$ | APASS |
| $r$ (mag) | $12.813 \pm 0.030$ | $12.375 \pm 0.040$ | $14.015 \pm 0.020$ | $13.551 \pm 0.030$ | APASS |
| $i$ (mag) | $12.763 \pm 0.060$ | $12.14 \pm 0.11$ | $13.890 \pm 0.030$ | $13.428 \pm 0.080$ | APASS |
| $J$ (mag) | $11.968 \pm 0.024$ | $11.405 \pm 0.023$ | $13.083 \pm 0.023$ | $12.638 \pm 0.026$ | 2MASS |
| $H$ (mag) | $11.780 \pm 0.026$ | $11.145 \pm 0.025$ | $12.827 \pm 0.023$ | $12.346 \pm 0.024$ | 2MASS |
| $K_s$ (mag) | $11.705 \pm 0.021$ | $11.095 \pm 0.023$ | $12.761 \pm 0.026$ | $12.327 \pm 0.026$ | 2MASS |

Note— Notes as for Table 8.

**Table 10.** Astrometric, Spectroscopic and Photometric parameters for HATS-68 and HATS-69

| Parameter | HATS-68 Value | HATS-69 Value | Source |
|---|---|---|---|
| Astrometric properties and cross-identifications | | | |
| 2MASS-ID | 01000141-5854172 | 19171138-6053301 | |
| TIC-ID | 322307342 | 467971286 | |
| GAIA DR2-ID | 4904279261014267648 | 6445881974332225536 | |
| R.A. (J2000) | $01^{\mathrm{h}}00^{\mathrm{m}}01.4134\mathrm{s}$ | $19^{\mathrm{h}}17^{\mathrm{m}}11.3641\mathrm{s}$ | GAIA DR2 |
| Dec. (J2000) | $-58^{\circ}54'17.1247''$ | $-60^{\circ}53'30.0584''$ | GAIA DR2 |
| $\mu_{\mathrm{R.A.}}$ (mas yr$^{-1}$) | $22.522 \pm 0.049$ | $8.699 \pm 0.027$ | GAIA DR2 |
| $\mu_{\mathrm{Dec.}}$ (mas yr$^{-1}$) | $7.594 \pm 0.041$ | $-17.887 \pm 0.023$ | GAIA DR2 |
| parallax (mas) | $1.627 \pm 0.028$ | $2.384 \pm 0.020$ | GAIA DR2 |
| Spectroscopic properties | | | |
| $T_{\mathrm{eff}\star}$ (K) | $6300 \pm 110$ | $5276 \pm 59$ | ZASPE |
| [Fe/H] | $0.180 \pm 0.057$ | $0.350 \pm 0.035$ | ZASPE |
| $v \sin i$ (km s$^{-1}$) | $7.42 \pm 0.29$ | $2.55 \pm 0.90$ | ZASPE |
| $v_{\mathrm{mac}}$ (km s$^{-1}$) | $4.79 \pm 0.16$ | $3.220 \pm 0.090$ | Assumed |
| $v_{\mathrm{mic}}$ (km s$^{-1}$) | $1.51 \pm 0.12$ | $0.831 \pm 0.027$ | Assumed |
| $\gamma_{\mathrm{RV}}$ (m s$^{-1}$) | $11894.6 \pm 5.9$ | $4087 \pm 29$ | FEROS |
| Photometric properties | | | |
| $G$ (mag) | $12.16310 \pm 0.00020$ | $13.76430 \pm 0.00020$ | GAIA DR2 |
| $BP$ (mag) | $12.4545 \pm 0.0017$ | $14.2527 \pm 0.0011$ | GAIA DR2 |
| $RP$ (mag) | $11.72880 \pm 0.00090$ | $13.13730 \pm 0.00090$ | GAIA DR2 |
| $B$ (mag) | $12.799 \pm 0.010$ | $14.916 \pm 0.020$ | APASS |
| $V$ (mag) | $12.276 \pm 0.020$ | $13.945 \pm 0.010$ | APASS |
| $g$ (mag) | $12.484 \pm 0.020$ | $14.401 \pm 0.030$ | APASS |
| $r$ (mag) | $12.137 \pm 0.010$ | $13.622 \pm 0.020$ | APASS |
| $i$ (mag) | $12.050 \pm 0.030$ | $13.598 \pm 0.030$ | APASS |
| $J$ (mag) | $11.250 \pm 0.026$ | $12.413 \pm 0.024$ | 2MASS |
| $H$ (mag) | $10.985 \pm 0.024$ | $11.968 \pm 0.025$ | 2MASS |
| $K_s$ (mag) | $10.949 \pm 0.019$ | $11.875 \pm 0.023$ | 2MASS |

Note— Notes as for Table 8.

**Table 11.** Derived stellar parameters for HATS-60, HATS-61, HATS-62 and HATS-63

| Parameter | **HATS-60** Value | **HATS-61** Value | **HATS-62** Value | **HATS-63** Value |
|---|---|---|---|---|
| $M_\star$ ($M_\odot$) ........ | $1.097^{+0.010}_{-0.016}$ | $1.076 \pm 0.014$ | $0.896^{+0.015}_{-0.010}$ | $0.931 \pm 0.019$ |
| $R_\star$ ($R_\odot$) ......... | $1.460 \pm 0.024$ | $1.664 \pm 0.024$ | $0.933^{+0.019}_{-0.013}$ | $1.070 \pm 0.012$ |
| $\log g_\star$ (cgs) ....... | $4.148 \pm 0.014$ | $4.028 \pm 0.012$ | $4.451 \pm 0.019$ | $4.349 \pm 0.015$ |
| $\rho_\star$ ($\mathrm{g\,cm^{-3}}$) ...... | $0.496 \pm 0.023$ | $0.330 \pm 0.014$ | $1.556 \pm 0.095$ | $1.071 \pm 0.047$ |
| $L_\star$ ($L_\odot$).......... | $1.996 \pm 0.066$ | $2.340 \pm 0.063$ | $0.671^{+0.029}_{-0.019}$ | $1.028 \pm 0.022$ |
| $T_{\mathrm{eff}\star}$ (K) ......... | $5688 \pm 20$ | $5542 \pm 21$ | $5416^{+19}_{-13}$ | $5627 \pm 18$ |
| [Fe/H]........... | $0.335 \pm 0.028$ | $0.247 \pm 0.037$ | $0.133 \pm 0.023$ | $0.081 \pm 0.038$ |
| Age (Gyr)........ | $7.55^{+0.70}_{-0.30}$ | $8.90^{+0.31}_{-0.41}$ | $9.55^{+0.99}_{-1.55}$ | $10.3 \pm 1.1$ |
| $A_V$ (mag)........ | $0.156 \pm 0.014$ | $0.137 \pm 0.013$ | $0.170 \pm 0.011$ | $0.081 \pm 0.011$ |
| Distance (pc)..... | $494.3 \pm 7.8$ | $694.0 \pm 8.8$ | $516.8^{+10.9}_{-6.1}$ | $634.8 \pm 6.2$ |

Note— The listed parameters are those determined through the joint differential evolution Markov Chain analysis described in Section 3.2. For all four systems the fixed-circular-orbit model has a higher Bayesian evidence than the eccentric-orbit model. We therefore assume a fixed circular orbit in generating the parameters listed here.

**Table 12.** Derived stellar parameters for HATS-64, HATS-65, HATS-66 and HATS-67

| Parameter | **HATS-64** Value | **HATS-65** Value | **HATS-66** Value | **HATS-67** Value |
|---|---|---|---|---|
| $M_\star$ ($M_\odot$) ........ | $1.564 \pm 0.028$ | $1.257 \pm 0.028$ | $1.411 \pm 0.022$ | $1.435 \pm 0.021$ |
| $R_\star$ ($R_\odot$) ......... | $2.113 \pm 0.071$ | $1.310 \pm 0.027$ | $1.841 \pm 0.041$ | $1.441 \pm 0.026$ |
| $\log g_\star$ (cgs) ....... | $3.982 \pm 0.024$ | $4.303 \pm 0.020$ | $4.057 \pm 0.017$ | $4.278 \pm 0.015$ |
| $\rho_\star$ ($\mathrm{g\,cm^{-3}}$) ...... | $0.234 \pm 0.020$ | $0.788 \pm 0.052$ | $0.318 \pm 0.019$ | $0.677 \pm 0.035$ |
| $L_\star$ ($L_\odot$).......... | $7.37 \pm 0.52$ | $2.38 \pm 0.11$ | $5.85 \pm 0.28$ | $3.52 \pm 0.16$ |
| $T_{\mathrm{eff}\star}$ (K) ......... | $6554 \pm 27$ | $6277 \pm 30$ | $6626 \pm 35$ | $6594 \pm 33$ |
| [Fe/H]........... | $0.220 \pm 0.039$ | $0.199 \pm 0.055$ | $-0.017 \pm 0.043$ | $0.332 \pm 0.052$ |
| Age (Gyr)........ | $1.861^{+0.097}_{-0.180}$ | $1.78 \pm 0.55$ | $2.17^{+0.16}_{-0.11}$ | $0.51 \pm 0.24$ |
| $A_V$ (mag)........ | $0.230 \pm 0.014$ | $0.243 \pm 0.014$ | $0.390 \pm 0.019$ | $0.385 \pm 0.018$ |
| Distance (pc)..... | $1083 \pm 36$ | $495 \pm 10$ | $1538 \pm 34$ | $982 \pm 19$ |

Note— Notes as for Table 11.

**Table 13.** Derived stellar parameters for HATS-68 and HATS-69

| | **HATS-68** | **HATS-69** |
|---|---|---|
| Parameter | Value | Value |
| $M_\star$ ($M_\odot$) ........ | $1.351 \pm 0.014$ | $0.892^{+0.011}_{-0.016}$ |
| $R_\star$ ($R_\odot$) ......... | $1.748 \pm 0.026$ | $0.8785 \pm 0.0077$ |
| $\log g_\star$ (cgs) ....... | $4.083 \pm 0.012$ | $4.501 \pm 0.013$ |
| $\rho_\star$ ($\mathrm{g\,cm^{-3}}$) ...... | $0.356 \pm 0.015$ | $1.854 \pm 0.070$ |
| $L_\star$ ($L_\odot$) .......... | $3.91 \pm 0.13$ | $0.4813 \pm 0.0084$ |
| $T_{\rm eff\star}$ (K) ......... | $6147 \pm 22$ | $5137 \pm 16$ |
| [Fe/H] ............ | $0.210 \pm 0.043$ | $0.377 \pm 0.034$ |
| Age (Gyr) ........ | $3.02 \pm 0.11$ | $8.0^{+1.8}_{-1.3}$ |
| $A_V$ (mag) ........ | $0.062 \pm 0.012$ | $0.155 \pm 0.012$ |
| Distance (pc) ..... | $618.2 \pm 9.3$ | $420.3 \pm 3.2$ |

NOTE— Notes as for Table 11.

**Table 14**. Orbital and planetary parameters for HATS-60b–HATS-64b

| Parameter | HATS-60b Value | HATS-61b Value | HATS-62b Value | HATS-63b Value |
|---|---|---|---|---|
| Light curve parameters | | | | |
| $P$ (days) | $3.560829 \pm 0.000032$ | $7.817953 \pm 0.000024$ | $3.2768837 \pm 0.0000033$ | $3.0566527 \pm 0.0000049$ |
| $T_c$ (BJD) [a] | $2458015.72358 \pm 0.00085$ | $2457673.0611 \pm 0.0014$ | $2455808.05158 \pm 0.00043$ | $2457659.93755 \pm 0.00089$ |
| $T_{14}$ (days) [a] | $0.1608 \pm 0.0019$ | $0.2304 \pm 0.0029$ | $0.11522 \pm 0.00093$ | $0.1020 \pm 0.0017$ |
| $T_{12} = T_{34}$ (days) [a] | $0.01485 \pm 0.00083$ | $0.0209 \pm 0.0013$ | $0.01348 \pm 0.00060$ | $0.0206 \pm 0.0012$ |
| $a/R_\star$ | $6.93 \pm 0.11$ | $10.23 \pm 0.14$ | $9.59 \pm 0.19$ | $8.09 \pm 0.12$ |
| $\zeta/R_\star$ [b] | $13.69 \pm 0.17$ | $9.54 \pm 0.13$ | $19.64^{+0.11}_{-0.14}$ | $24.14 \pm 0.61$ |
| $R_p/R_\star$ | $0.0811 \pm 0.0033$ | $0.0738 \pm 0.0040$ | $0.1159 \pm 0.0011$ | $0.1159 \pm 0.0032$ |
| $b^2$ | $0.202^{+0.032}_{-0.035}$ | $0.258^{+0.026}_{-0.027}$ | $0.121^{+0.036}_{-0.032}$ | $0.525^{+0.030}_{-0.028}$ |
| $b \equiv a \cos i/R_\star$ | $0.450^{+0.034}_{-0.041}$ | $0.508^{+0.025}_{-0.027}$ | $0.348^{+0.048}_{-0.049}$ | $0.724^{+0.020}_{-0.020}$ |
| $i$ (deg) | $86.28 \pm 0.35$ | $87.15 \pm 0.18$ | $87.92 \pm 0.35$ | $84.86 \pm 0.19$ |
| HATSouth dilution factors [c] | | | | |
| Dilution factor 1 | $0.788 \pm 0.082$ | $0.796 \pm 0.098$ | $0.962 \pm 0.025$ | $0.962 \pm 0.039$ |
| Dilution factor 2 | $0.494 \pm 0.089$ | ··· | ··· | ··· |
| Limb-darkening coefficients [d] | | | | |
| $c_1, r$ | 0.3914 | 0.3961 | 0.4133 | 0.3863 |
| $c_2, r$ | 0.3109 | 0.3057 | 0.2925 | 0.3084 |
| $c_1, R$ | ··· | ··· | 0.3853 | 0.3601 |
| $c_2, R$ | ··· | ··· | 0.2977 | 0.3120 |
| $c_1, i$ | ··· | 0.2954 | 0.3127 | 0.2917 |
| $c_2, i$ | ··· | 0.3220 | 0.3078 | 0.3181 |
| $c_1, z$ | ··· | ··· | 0.2427 | ··· |
| $c_2, z$ | ··· | ··· | 0.3129 | ··· |
| RV parameters | | | | |
| $K$ ($\mathrm{m\,s^{-1}}$) | $82.6 \pm 6.9$ | $330 \pm 13$ | $10.2 \pm 7.8$ | $140 \pm 18$ |
| $e$ [e] | $< 0.191$ | $< 0.092$ | $< 0.298$ | $< 0.136$ |
| RV jitter FEROS ($\mathrm{m\,s^{-1}}$) [f] | $16.9 \pm 5.7$ | $26 \pm 11$ | $56 \pm 14$ | $43 \pm 10$ |
| RV jitter HARPS ($\mathrm{m\,s^{-1}}$) | $< 10.0$ | ··· | $34 \pm 15$ | ··· |
| RV jitter Coralie ($\mathrm{m\,s^{-1}}$) | ··· | $67 \pm 55$ | $2.0 \pm 1.7$ | ··· |
| RV jitter PFS ($\mathrm{m\,s^{-1}}$) | ··· | ··· | $34 \pm 15$ | ··· |
| RV jitter CYCLOPS ($\mathrm{m\,s^{-1}}$) | ··· | ··· | $1 \pm 29$ | ··· |
| Planetary parameters | | | | |
| $M_p$ ($M_J$) | $0.662 \pm 0.055$ | $3.40 \pm 0.14$ | $< 0.179$ | $0.96 \pm 0.12$ |
| $R_p$ ($R_J$) | $1.153 \pm 0.053$ | $1.195 \pm 0.067$ | $1.055 \pm 0.025$ | $1.207 \pm 0.039$ |
| $C(M_p, R_p)$ [g] | 0.14 | −0.00 | 0.39 | −0.03 |
| $\rho_p$ ($\mathrm{g\,cm^{-3}}$) | $0.537^{+0.100}_{-0.070}$ | $2.47 \pm 0.44$ | $0.076 \pm 0.053$ | $0.67 \pm 0.11$ |
| $\log g_p$ (cgs) | $3.093 \pm 0.050$ | $3.770 \pm 0.052$ | $2.20^{+0.25}_{-0.49}$ | $3.211 \pm 0.065$ |
| $a$ (AU) | $0.04708^{+0.00015}_{-0.00023}$ | $0.07908 \pm 0.00033$ | $0.04163^{+0.00024}_{-0.00016}$ | $0.04026 \pm 0.00028$ |
| $T_{eq}$ (K) | $1528 \pm 11$ | $1226.1 \pm 7.3$ | $1237 \pm 12$ | $1398.3 \pm 9.0$ |
| $\Theta$ [h] | $0.0493 \pm 0.0044$ | $0.415 \pm 0.029$ | $0.0062 \pm 0.0046$ | $0.0681 \pm 0.0090$ |
| $\log_{10}\langle F \rangle$ (cgs) [i] | $9.090 \pm 0.013$ | $8.707 \pm 0.010$ | $8.722^{+0.019}_{-0.014}$ | $8.936 \pm 0.011$ |

*Table 14 continued*

**Table 14** *(continued)*

| Parameter | HATS-60b Value | HATS-61b Value | HATS-62b Value | HATS-63b Value |
|---|---|---|---|---|

Note— For all systems we adopt a model in which the orbit is assumed to be circular. See the discussion in Section 3.2.

[a] Times are in Barycentric Julian Date calculated directly from UTC *without* correction for leap seconds. $T_C$: Reference epoch of mid transit that minimizes the correlation with the orbital period. $T_{12}$: total transit duration, time between first to last contact; $T_{12} = T_{34}$: ingress/egress time, time between first and second, or third and fourth contact.

[b] Reciprocal of the half duration of the transit used as a jump parameter in our MCMC analysis in place of $a/R_\star$. It is related to $a/R_\star$ by the expression $\zeta/R_\star = a/R_\star(2\pi(1 + e\sin\omega))/(P\sqrt{1-b^2}\sqrt{1-e^2})$ (Bakos et al. 2010).

[c] Scaling factor applied to the model transit that is fit to the HATSouth light curves. This factor accounts for dilution of the transit due to blending from neighboring stars and over-filtering of the light curve. These factors are varied in the fit, with independent values adopted for each HATSouth light curve. The factors listed for HATS-61, HATS-62 and HATS-63 are for the G548.4, G582.1 and G597.2 light curves, respectively. For HATS-60 we list the factors for the G537.3 and G537.4 light curves in order.

[d] Values for a quadratic law, adopted from the tabulations by Claret (2004) according to the spectroscopic (ZASPE) parameters listed in Table 8.

[e] The 95% confidence upper limit on the eccentricity determined when $\sqrt{e}\cos\omega$ and $\sqrt{e}\sin\omega$ are allowed to vary in the fit.

[f] Term added in quadrature to the formal RV uncertainties for each instrument. This is treated as a free parameter in the fitting routine. In cases where the jitter is consistent with zero, we list its 95% confidence upper limit.

[g] Correlation coefficient between the planetary mass $M_p$ and radius $R_p$ estimated from the posterior parameter distribution.

[h] The Safronov number is given by $\Theta = \frac{1}{2}(V_{\rm esc}/V_{\rm orb})^2 = (a/R_p)(M_p/M_\star)$ (see Hansen & Barman 2007).

[i] Incoming flux per unit surface area, averaged over the orbit.

**Table 15**. Orbital and planetary parameters for HATS-64b–HATS-67b

| Parameter | HATS-64b Value | HATS-65b Value | HATS-66b Value | HATS-67b Value |
|---|---|---|---|---|
| Light curve parameters | | | | |
| $P$ (days) | $4.908897 \pm 0.000013$ | $3.1051610 \pm 0.0000016$ | $3.1414391 \pm 0.0000074$ | $1.6091788 \pm 0.0000040$ |
| $T_c$ (BJD) | $2457769.82287 \pm 0.00082$ | $2457520.96130 \pm 0.00041$ | $2457603.0514 \pm 0.0014$ | $2457796.88127 \pm 0.00043$ |
| $T_{14}$ (days) | $0.2419 \pm 0.0020$ | $0.1202 \pm 0.0013$ | $0.1893 \pm 0.0027$ | $0.0811 \pm 0.0011$ |
| $T_{12} = T_{34}$ (days) | $0.0202 \pm 0.0014$ | $0.0214 \pm 0.0012$ | $0.0149 \pm 0.0011$ | $0.0303 \pm 0.0054$ |
| $a/R_\star$ | $6.68 \pm 0.20$ | $7.38 \pm 0.16$ | $5.50 \pm 0.11$ | $4.526 \pm 0.077$ |
| $\zeta/R_\star$ | $9.026 \pm 0.063$ | $20.02 \pm 0.16$ | $11.47 \pm 0.18$ | $34.77 \pm 0.61$ |
| $R_p/R_\star$ | $0.0817 \pm 0.0024$ | $0.1181 \pm 0.0025$ | $0.0787 \pm 0.0043$ | $0.1201 \pm 0.0020$ |
| $b^2$ | $0.103^{+0.059}_{-0.054}$ | $0.445^{+0.024}_{-0.024}$ | $0.080^{+0.049}_{-0.045}$ | $0.742^{+0.011}_{-0.012}$ |
| $b \equiv a\cos i/R_\star$ | $0.321^{+0.082}_{-0.101}$ | $0.667^{+0.018}_{-0.019}$ | $0.283^{+0.076}_{-0.095}$ | $0.8613^{+0.0065}_{-0.0071}$ |
| $i$ (deg) | $87.24 \pm 0.85$ | $84.82 \pm 0.26$ | $87.06 \pm 0.92$ | $79.03 \pm 0.26$ |
| HATSouth dilution factors | | | | |
| Dilution factor 1 | $0.658 \pm 0.066$ | $0.755 \pm 0.039$ | $0.608 \pm 0.098$ | $0.973 \pm 0.030$ |
| Dilution factor 2 | ··· | ··· | ··· | $0.934 \pm 0.039$ |
| Limb-darkening coefficients | | | | |
| $c_1, r$ | 0.2300 | 0.2032 | 0.2432 | 0.2437 |
| $c_2, r$ | 0.3919 | 0.3475 | 0.3853 | 0.4008 |
| $c_1, R$ | 0.2087 | 0.2104 | ··· | ··· |
| $c_2, R$ | 0.3913 | 0.3885 | ··· | ··· |
| $c_1, i$ | 0.1561 | 0.1394 | 0.1693 | 0.1660 |
| $c_2, i$ | 0.3827 | 0.3395 | 0.3766 | 0.3962 |
| RV parameters | | | | |
| $K$ (m s$^{-1}$) | $85 \pm 18$ | $97.7 \pm 9.9$ | $586 \pm 75$ | $194 \pm 16$ |
| $e$ | $< 0.151$ | $< 0.062$ | $< 0.064$ | $< 0.057$ |
| RV jitter FEROS (m s$^{-1}$) | $62 \pm 14$ | $< 30.7$ | $< 259.3$ | $37 \pm 11$ |
| RV jitter HARPS (m s$^{-1}$) | $78 \pm 28$ | $< 2.7$ | ··· | ··· |
| Planetary parameters | | | | |
| $M_p$ ($M_{\rm J}$) | $0.96 \pm 0.20$ | $0.821 \pm 0.083$ | $5.33 \pm 0.68$ | $1.45 \pm 0.12$ |
| $R_p$ ($R_{\rm J}$) | $1.679 \pm 0.081$ | $1.501 \pm 0.050$ | $1.411 \pm 0.084$ | $1.685 \pm 0.047$ |
| $C(M_p, R_p)$ | 0.11 | 0.15 | 0.06 | −0.12 |
| $\rho_p$ (g cm$^{-3}$) | $0.245^{+0.068}_{-0.050}$ | $0.300 \pm 0.039$ | $2.34^{+0.56}_{-0.43}$ | $0.374 \pm 0.047$ |

*Table 15 continued*

**Table 15** *(continued)*

| Parameter | HATS-64b Value | HATS-65b Value | HATS-66b Value | HATS-67b Value |
|---|---|---|---|---|
| $\log g_p$ (cgs) | $2.92 \pm 0.10$ | $2.953 \pm 0.050$ | $3.820 \pm 0.075$ | $3.101 \pm 0.045$ |
| $a$ (AU) | $0.06562 \pm 0.00039$ | $0.04497 \pm 0.00033$ | $0.04714 \pm 0.00025$ | $0.03032 \pm 0.00015$ |
| $T_{\rm eq}$ (K) | $1793 \pm 27$ | $1634 \pm 18$ | $1998 \pm 21$ | $2193 \pm 22$ |
| $\Theta$ | $0.048 \pm 0.010$ | $0.0390 \pm 0.0040$ | $0.251 \pm 0.035$ | $0.0355 \pm 0.0032$ |
| $\log_{10}\langle F \rangle$ (cgs) | $9.367 \pm 0.026$ | $9.205 \pm 0.019$ | $9.555 \pm 0.019$ | $9.716 \pm 0.017$ |

Note— Notes as for Table 14. The HATSouth dilution factors listed for HATS-64, HATS-65, and HATS-66 are for the G606.3, G625.2, and G601.1 light curves, respectively. For HATS-67 we list the factors for the G698.1 and G698.4 light curves in order.

**Table 16**. Orbital and planetary parameters for HATS-68b and HATS-69b

| Parameter | HATS-68b Value | HATS-69b Value |
|---|---|---|
| Light curve parameters | | |
| $P$ (days) | $3.5862202 \pm 0.0000047$ | $2.2252577 \pm 0.0000019$ |
| $T_c$ (BJD) | $2457410.4086 \pm 0.0011$ | $2457755.39390 \pm 0.00052$ |
| $T_{14}$ (days) | $0.1425 \pm 0.0017$ | $0.09824 \pm 0.00091$ |
| $T_{12} = T_{34}$ (days) | $0.01976 \pm 0.00085$ | $0.01019 \pm 0.00036$ |
| $a/R_\star$ | $6.232 \pm 0.086$ | $7.859 \pm 0.099$ |
| $\zeta/R_\star$ | $16.16^{+0.28}_{-0.21}$ | $22.71 \pm 0.23$ |
| $R_p/R_\star$ | $0.0725 \pm 0.0016$ | $0.1105 \pm 0.0023$ |
| $b^2$ | $0.542^{+0.021}_{-0.017}$ | $0.043^{+0.034}_{-0.027}$ |
| $b \equiv a \cos i/R_\star$ | $0.736^{+0.014}_{-0.012}$ | $0.207^{+0.071}_{-0.081}$ |
| $i$ (deg) | $83.21 \pm 0.19$ | $88.49 \pm 0.55$ |
| HATSouth dilution factors | | |
| Dilution factor 1 | $0.9937 \pm 0.0031$ | $0.890 \pm 0.049$ |
| Limb-darkening coefficients | | |
| $c_1, g$ | ··· | 0.7272 |
| $c_2, g$ | ··· | 0.1014 |
| $c_1, r$ | 0.2706 | 0.4905 |
| $c_2, r$ | 0.3797 | 0.2449 |
| $c_1, i$ | 0.1913 | 0.3701 |
| $c_2, i$ | 0.3762 | 0.2786 |
| RV parameters | | |
| $K$ (m s$^{-1}$) | $139.3 \pm 6.4$ | $52 \pm 28$ |
| $e$ | $< 0.036$ | $< 0.519$ |
| RV jitter FEROS (m s$^{-1}$) | $< 4.4$ | $117 \pm 21$ |
| RV jitter HARPS (m s$^{-1}$) | $< 2.0$ | ··· |
| RV jitter Coralie (m s$^{-1}$) | $81 \pm 34$ | ··· |
| Planetary parameters | | |
| $M_p$ ($M_{\rm J}$) | $1.290 \pm 0.059$ | $< 0.577$ |
| $R_p$ ($R_{\rm J}$) | $1.232^{+0.039}_{-0.029}$ | $0.945 \pm 0.022$ |
| $C(M_p, R_p)$ | 0.04 | −0.03 |
| $\rho_p$ (g cm$^{-3}$) | $0.856 \pm 0.083$ | $0.46 \pm 0.25$ |
| $\log g_p$ (cgs) | $3.325 \pm 0.031$ | $2.94^{+0.18}_{-0.42}$ |
| $a$ (AU) | $0.05071 \pm 0.00018$ | $0.03211^{+0.00014}_{-0.00019}$ |
| $T_{\rm eq}$ (K) | $1741 \pm 12$ | $1295.7 \pm 6.9$ |
| $\Theta$ | $0.0782 \pm 0.0042$ | $0.024 \pm 0.013$ |
| $\log_{10}\langle F \rangle$ (cgs) | $9.316 \pm 0.012$ | $8.8028 \pm 0.0092$ |

Note— Notes as for Table 14. The HATSouth dilution factors listed for HATS-68 and HATS-69 are for the G755.3 and G778.4 light curves, respectively.

**Table 17**. Comparison Between Isochrone and Empirical Model Results for Stellar Mass and Radius

| System | Isoc. $M_\star$ $M_\odot$ | Empir. $M_\star$ $M_\odot$ | Isoc. $R_\star$ $R_\odot$ | Empir. $R_\star$ $R_\odot$ |
|---|---|---|---|---|
| HATS-60 | $1.097^{+0.010}_{-0.016}$ | $0.91^{+0.23}_{-0.30}$ | $1.460 \pm 0.024$ | $1.448 \pm 0.028$ |
| HATS-61 | $1.076 \pm 0.014$ | $0.99^{+0.16}_{-0.24}$ | $1.664 \pm 0.024$ | $1.649 \pm 0.025$ |
| HATS-62 | $0.896^{+0.015}_{-0.010}$ | $1.06 \pm 0.11$ | $0.933^{+0.019}_{-0.013}$ | $0.956 \pm 0.023$ |
| HATS-63 | $0.931 \pm 0.019$ | $0.86 \pm 0.20$ | $1.070 \pm 0.012$ | $1.066 \pm 0.015$ |
| HATS-64 | $1.564 \pm 0.028$ | $1.60^{+0.22}_{-0.15}$ | $2.113 \pm 0.071$ | $2.136 \pm 0.071$ |
| HATS-65 | $1.257 \pm 0.028$ | $1.62 \pm 0.13$ | $1.310 \pm 0.027$ | $1.320 \pm 0.029$ |
| HATS-66 | $1.411 \pm 0.022$ | $1.27 \pm 0.27$ | $1.841 \pm 0.041$ | $1.850 \pm 0.049$ |
| HATS-67 | $1.435 \pm 0.021$ | $1.47^{+0.14}_{-0.19}$ | $1.441 \pm 0.026$ | $1.431 \pm 0.042$ |
| HATS-68 | $1.351 \pm 0.014$ | $0.87^{+0.21}_{-0.16}$ | $1.748 \pm 0.026$ | $1.764 \pm 0.067$ |
| HATS-69 | $0.892^{+0.011}_{-0.016}$ | $0.747^{+0.097}_{-0.055}$ | $0.8785 \pm 0.0077$ | $0.864 \pm 0.018$ |

**Table 18.** Blend Analysis Results

| System | H,S-s | | H,S-s$_{\rm BGEB}$ | | H-p,s | | $M_3$[a] | $M_3$ 95% U.L.[b] | $L_3/L_1$ 95% U.L.[c] |
|---|---|---|---|---|---|---|---|---|---|
| | $\Delta\chi^2_{\rm tot}$ | $\Delta\chi^2_{\varpi}$ | $\Delta\chi^2_{\rm tot}$ | $\Delta\chi^2_{\varpi}$ | $\Delta\chi^2_{\rm tot}$ | $\Delta\chi^2_{\varpi}$ | $M_\odot$ | $M_\odot$ | |
| HATS-60 | 158.8 | 138.8 | 10.8 | 10.1 | 0.8 | 0.0 | 0.20 | 0.42 | 0.0082 |
| HATS-61 | 363.2 | 59.5 | 26.4 | −7.0 | −2.9 | −7.0 | 0.82 | 0.85 | 0.2038 |
| HATS-62 | 160.7 | 156.4 | 61.2 | 12.1 | −16.3 | −4.8 | 0.45 | 0.56 | 0.0760 |
| HATS-63 | 87.2 | 7.3 | 3.2 | 0.0 | 0.3 | −0.1 | 0.23 | 0.35 | 0.0116 |
| HATS-64 | 67.5 | 53.1 | 58.0 | 50.9 | −8.1 | −0.5 | 1.01 | 1.32 | 0.4220 |
| HATS-65 | 38.4 | 54.1 | 33.3 | 48.5 | −11.4 | −0.1 | 0.53 | 0.67 | 0.0407 |
| HATS-66 | 126.5 | 97.8 | 49.0 | 18.5 | −0.6 | 0.0 | 0.37 | 0.66 | 0.0180 |
| HATS-67 | 259.9 | 47.9 | 55.3 | 40.0 | −1.0 | 0.0 | 0.33 | 0.50 | 0.0072 |
| HATS-68 | 18.6 | 0.0 | 9.9 | 2.3 | −0.7 | −0.1 | 0.20 | 0.78 | 0.0565 |
| HATS-69 | 589.2 | 64.0 | 54.9 | 6.0 | 1.2 | −0.2 | 0.21 | 0.28 | 0.0131 |

Note— We follow the convention of Hartman et al. (2009) in referring to different blend scenarios. H,S-s corresponds to a hierarchical triple star system where the two fainter sources form an eclipsing binary. H,S-s$_{BGEB}$ is a blend between a bright foreground star and a fainter background eclipsing binary star system. H-p,s is a bright star with a transiting planet and a fainter unresolved stellar companion. The $\Delta\chi^2$ values provided are the difference in $\chi^2$ between the best-fit model of each type and the best-fit H-p model, corresponding to a single star with a transiting planet. We list both the total $\Delta\chi^2$, and the contribution to $\Delta\chi^2$ from the parallax $\varpi$. The rejection of blended eclipsing binary scenarios for each of these systems is discussed in Section 3.5.

[a] The mass of the contaminating unresolved stellar companion for the best-fit H-p,s model.

[b] 95% ($\Delta\chi^2 = 3.84$) confidence upper limit on the mass of the contaminating unresolved stellar companion for the H-p,s model.

[c] 95% ($\Delta\chi^2 = 3.84$) confidence upper limit on the ratio of the luminosity of the contaminating unresolved stellar companion to the luminosity of the planet host for the H-p,s model.

**Table 19.** Relative radial velocities and bisector spans for HATS-60–HATS-69.

| System | BJD | RV[a] | $\sigma_{\rm RV}$[b] | BS | $\sigma_{\rm BS}$ | Phase | Instrument |
|---|---|---|---|---|---|---|---|
| | (2,450,000+) | (m s$^{-1}$) | (m s$^{-1}$) | (m s$^{-1}$) | (m s$^{-1}$) | | |
| HATS-60 | 7866.89844 | −71.80 | 15.60 | 37.0 | 21.0 | 0.205 | HARPS |
| HATS-60 | 7867.92159 | −25.00 | 17.70 | −29.0 | 23.0 | 0.493 | HARPS |
| HATS-60 | 7868.88460 | 36.50 | 34.20 | 60.0 | 45.0 | 0.763 | HARPS |
| HATS-60 | 7870.88835 | −71.70 | 10.30 | 1.0 | 13.0 | 0.326 | HARPS |
| HATS-60 | 7871.90927 | 73.50 | 10.40 | 21.0 | 13.0 | 0.613 | HARPS |
| HATS-60 | 7910.85820 | 24.95 | 9.10 | −27.0 | 13.0 | 0.551 | FEROS |
| HATS-60 | 7911.83630 | 65.75 | 11.80 | −41.0 | 17.0 | 0.825 | FEROS |
| HATS-60 | 7914.78684 | 64.15 | 9.80 | 12.0 | 14.0 | 0.654 | FEROS |
| HATS-60 | 7915.82677 | 34.05 | 7.90 | −36.0 | 12.0 | 0.946 | FEROS |
| HATS-60 | 7967.78362 | 58.35 | 10.60 | −27.0 | 15.0 | 0.537 | FEROS |

[a] The zero-point of these velocities is arbitrary. An overall offset $\gamma_{\rm rel}$ fitted independently to the velocities from each instrument has been subtracted.

[b] Internal errors excluding the component of astrophysical jitter considered in Section 3.2.

Note— This table is available in a machine-readable form in the online journal. A portion is shown here for guidance regarding its form and content.